\documentclass[12pt,a4paper]{iopart}
\usepackage{amssymb,graphicx}
\usepackage{iopams}
\usepackage{setstack,cite,braket,float}
\usepackage[hang,nooneline]{subfigure}
\newcounter{fig}

\newcommand{\bea}{\begin{eqnarray}}
\newcommand{\eea}{\end{eqnarray}}
\newcommand{\bes}{\numparts}
\newcommand{\ees}{\endnumparts}

\begin{document}
\bibliography{revtex4}
\title[Non-autonomous bright solitons and their stability in Rabi coupled binary BECs]{Non-autonomous bright solitons and their stability in Rabi coupled binary Bose-Einstein Condensates}
\author{T Kanna$^\dag$, R Babu Mareeswaran$^\dag$, and Franz G Mertens$^\ddag$}
\address{$^\dag$Post Graduate and Research Department of Physics, Bishop Heber College, Tiruchirappalli--620 017, Tamil Nadu, India}
\address{$^\ddag$Physikalisches Institut, Universit\"at Bayreuth, D-95440 Bayreuth, Germany}
\eads{kanna\_phy@bhc.edu.in (corresponding author), franzmertens@gmail.com}

\begin{abstract}
The dynamics of non-autonomous bright matter-wave solitons in Rabi coupled binary
Bose-Einstein condensates is explored. By performing a unitary and similarity/lens-type transformation,
 we reduce the non-autonomous Gross-Pitaevskii equation into the celebrated Manakov model.
 Then, we construct the exact bright solitons of the non-autonomous Gross-Pitaevskii system in
 the presence of time dependent nonlinearities for two specific forms, namely hyperbolic nonlinearities,
 which are of physical interest. The experimental possibilities of realizing
the forms of temporally modulated potentials corresponding to these time dependent nonlinearities and
supporting such localized structures are explored. Our study on
the propagation of one soliton shows that the amplitude, velocity and shape of the bright soliton are altered by the
time-dependent
 scattering length. We also analyse the non-trivial energy sharing collision of Manakov solitons in the presence
 of Rabi coupling and
 aforementioned nonlinearities. We find
that breathers arise in two-soliton collisions and the nature of energy sharing collisions
is altered from that of Manakov system due to Rabi coupling only. Further in the presence of
 time-dependent nonlinearities the collision scenario is again altered significantly.
 Finally, the stability of these localized structures is examined
using a recently
developed powerful analytic method by Quintero et.~al., [Phys.~Rev.~E~91, 012905~(2015)] and it is shown that the
non-autonomous bright solitons are indeed stable. The evolution of position and velocity is also studied.
\end{abstract}
\pacs{ 05.45.Yv, 02.30.IK, 03.75.Lm, 03.75.-b}
\maketitle
\section{Introduction}
Bose-Einstein condensates (BECs) have become an
important ground for the study of macroscopic quantum
phenomena \cite{P.G}. Since the first realization
of BECs with alkali atoms, various nonlinear structures have been experimentally observed
and/or theoretically investigated, such as bright solitons \cite{bsol}, dark solitons \cite{dsol}, dark-bright
solitons \cite{dbsol}, vortices \cite{vortex}, Faraday waves \cite{fara}, skyrmions \cite{sky}, etc. BECs with tunable interatomic
interactions have been the subject of intense theoretical
and experimental interest in recent years \cite{fesh1}. In the vicinity of a Feshbach-resonance (FR),
the atomic scattering length depends sensitively on the applied external magnetic field (see refs.~\cite{corn,corn1}
and references therein), allowing the magnitude and sign of the atomic interactions to be tuned to any value. These
techniques offer some opportunities to achieve a nonlinearity
management through the use of time-dependent
and/or nonuniform fields. Utilizing this nonlinearity management concept, several nonlinear wave
patterns and effects, such as, non-autonomous bright solitons \cite{nonmag} as well as dark-bright
solitons \cite{babujpa}, Bloch oscillations \cite{bloch}, and rogue waves \cite{rogue} have been observed in BECs.

Multicomponent BECs are a mixture of different
atomic species (heteronulcear BEC mixtures) or a mixture of same species at different hyperfine states (homonuclear BEC
mixtures). Of the multi-component BECs, the simplest form is the two-component BEC.
In homonuclear BECs, in addition to the inter species interaction the two different
 hyperfine states can be coupled by ``Rabi coupling". This is a linear coupling between separate wave functions
 say $\ket{1}$ and $\ket{2}$ induced by radio-frequencies. When the coupling drive is turned on, suddenly,
 it will induce an extended
series of oscillations of the total population from
the $\ket{1}$ to the $\ket{2}$ state, so called ``Rabi oscillations" \cite{rabiexp}. Under the application of
Rabi coupling between the components
of a weakly interacting multicomponent BECs, one component
of BECs can be transferred to another \cite{rabidimitri,rabi1}.
Experiments have been performed for two-component $^{87}$Rb condensate
with atomic states customarily denoted by $\ket{1}$ and $\ket{2}$; in particular, these states can be either $\ket{F=2,m_F=1}$
and $\ket{2,2}$ \cite{rabiexp1}, or $\ket{1,-1}$ and $\ket{2,1}$ \cite{rabiexp}.
In a recent experimental work \cite{exp}, solitons have been observed in a binary Gross-Pitaevskii (GP) system with
 Rabi
coupling. Several other studies on Rabi coupled binary GP systems, namely magnetic solitons \cite{msol},
dark soliton \cite{dark}, domain walls \cite{domain}, vortex pairs \cite{vortexpair}, countersuperflow \cite{ushi},
and topological defects \cite{mason}, have also been investigated.

Motivated by these works, here we study theoretically
the non-autonomous bright solitons and their stability in a Rabi-coupled
quasi-one-dimensional GP system. In our earlier work, two of the authors (T.K. and R.B.) and co-workers
considered the repulsive condensates and studied the
dark-bright solitons dynamics. Now our attention is on attractive condensates and their soliton
patterns, which are distinctly different from those observed
in Ref.~\cite{babujpa}.

 On the other hand, stability of nonlinear waves in multi-component BECs is another critical issue in (1+1)-
 dimension as well as in higher dimensions. To the best of our knowledge so far no analytical tool has
 been developed to study the stability of multi-component non-autonomous soliton like structures appearing in
 integrable non-autonomous nonlinear evolution equations. Recently, Quintero et al.,\cite{stability} proposed an
 efficient method to determine
 the stability of bright solitons in both autonomous and non-autonomous settings of the nonlinear Schr\"odinger
 (NLS) equation.
 In this work, we apply that method to the obtained bright soliton solutions of the non-autonomous GP system
 (\ref{lngp1}) given below. The rest of the paper is organized as follows:

In section II, we describe the model equation for quasi one-dimensional Rabi-coupled BEC. In section III, we
show that under a suitable unitary and similarity type transformation, the Rabi-coupled non-autonomous GP system can
be converted into the famous integrable Manakov system along with a constraint condition.
In section IV, we will study the evolution of one- and two-solitons in autonomous and non-autonomous
Rabi-coupled GP systems and discuss their dynamical behavior for two particular physically interesting
forms of the time-dependent nonlinearity coefficient.
Following this, we address the stability of the obtained non-autonomous bright one-soliton by an
analytical procedure developed recently by Quintero et al.,\cite{stability}
and show that the non-autonomous soliton structures reported here are indeed stable.
Section VI contains the concluding remarks.
\section{Description of the Model}
 We consider a two-component BEC that is condensed
into two different hyperfine states $\ket{1}$ and $\ket{2}$ such as those of
$^{87}$Rb atoms \cite{rabiexp}. The two component BEC is assumed to be trapped in a
simple harmonic potential with the trapping in the transverse directions being
stronger. Then the BEC is cigar-shaped and is governed by the following
dimensionless one-dimensional (1D) GP equation \cite{P.G}:
\bes\label{lngp}\bea
i\frac{\partial \psi_1}{\partial t}=\left[-\frac{1}{2}\frac{\partial^2}{\partial x^2}+V_{\mbox{ext}}(x,t)\right]\psi_1+(g_{11}|\psi_1|^2+g_{12}|\psi_2|^2)\psi_1+\chi \psi_2, \\ \label{lngp11}
i\frac{\partial \psi_2}{\partial t}=\left[-\frac{1}{2}\frac{\partial^2}{\partial x^2}+V_{\mbox{ext}}(x,t)\right]\psi_2+ (g_{12}|\psi_1|^2+g_{22}|\psi_2|^2)\psi_2+\chi \psi_1. \label{lngp21}
\eea\ees
Here $\psi_j(x,t)(j=1,2)$ are the condensate wave functions in the two hyperfine states, spatial coordinate $x$ and
time $t$ are respectively measured in units of $a_0$ and $\omega_\bot^{-1}$, where $a_0=\sqrt{\hbar/m\omega_\bot}$($m$
denotes atomic mass) is the transverse harmonic oscillator length.
The coupling constants $g_{ii}=2a_{ii}/a_B$ for $i=1,2$ and $g_{12}=2a_{12}/a_B$ are
the intra-species and inter-species interaction strengths respectively, where $a_B$ is the Bohr radius and $a_{ij}$ are the $s$-wave scattering
 lengths of the species $i$ and $j$ respectively, and can be tuned with the aid of magnetic-field induced
 FR mechanism \cite{fesh,fesh3}; $V_{\mbox{ext}}(x,t)$ = $\frac{1}{2}\Omega^2(t)x^2$ (where $\Omega^2(t)$= $\omega_x^2(t)/2\omega_\bot^2$, in which $\omega_x$ and $\omega_\bot$ are the temporally modulated
axial trap frequency and radial frequency) is the time-dependent harmonic trap potential.
 The cross coupling term $\chi$ is the Rabi coupling parameter and is assumed to be real and positive.

Based on experimental results pertaining to two-component $^{87}$Rb BECs \cite{exptwo}, we
can assume that the scattering lengths to be equal, and also they can be tuned
through FR \cite{FR}. Hence, we consider the GP system (1) with  equal interaction strengths, i.e., $g_{11}=g_{12}=g_{22}=-\beta(t)$, where $\beta(t)$ is a
positive function. The resulting equations can be expressed as
\bes\bea
i \frac{\partial \psi_1}{\partial t}=\left[-\frac{1}{2}\frac{\partial^2}{\partial x^2}+V_{\mbox{ext}}(x,t)\right]\psi_1- \beta(t)(|\psi_1|^2+|\psi_2|^2)\psi_1+\chi \psi_2, \\ \label{lngp1a}
i \frac{\partial \psi_2}{\partial t}=\left[-\frac{1}{2}\frac{\partial^2}{\partial x^2}+V_{\mbox{ext}}(x,t)\right]\psi_2- \beta(t)(|\psi_1|^2+|\psi_2|^2)\psi_2+\chi \psi_1. \label{lngp1b}
\eea\label{lngp1}\ees
Next, we derive the continuity equation from the time-dependent GP equation (\ref{lngp1}). For this purpose,
the coupled GP system (2a) and (\ref{lngp1b}) is multiplied by $\psi_1^*$ and $\psi_2^*$ respectively,
and the complex conjugate equations of the coupled GP system (2a) and (\ref{lngp1b}) are multiplied respectively
by $\psi_1$ and $\psi_2$. Combining the resulting equations suitably, we get
 \bea
\frac{\partial \rho}{\partial t}+\frac{\partial j}{\partial x}=0
\label{con}
\eea
where norm density $\rho=(|\psi_1|^2+\psi_2|^2)$, and
momentum current density $j= (i/2)\left[(\psi_1^*\psi_{1,x}-\mbox{c.c})+(\psi_2^*\psi_{2,x}-\mbox{c.c})\right]$.
Here c.c denotes complex conjugation. The norm $(N)$ of Eq.~(\ref{lngp1})
is given by
\bes\bea
\mbox{Norm}: N &= &\int_{-\infty}^{\infty}\rho~dx.\label{norm}
\eea
We also require the following physical quantities, namely field momentum and normalized
momentum for the stability analysis of the non-autonomous solitons.
\bea
\mbox{Field Momentum}: p = \int_{-\infty}^{\infty}~j(x,t)~dx, \\ \label{moment}
\mbox{Normalized Momentum}: P =  \frac{p}{N}. \label{norm_moment}
\eea\ees

\section{Transforming Rabi coupled non-autonomous GP system to the Manakov system}
In a recent work \cite{babujpa}, two of the authors (T.K. and R.B.) along with their co-workers have converted Eq.~(\ref{lngp1}) with
repulsive nonlinearity to the defocusing coupled NLS system, a known integrable system by employing two successive
transformations. Here we consider Eq.~(\ref{lngp1}) with focusing nonlinearity, i.e., $\beta(t)>0$.
First, we utilize the following unitary transformation \cite{rabikev}
\bea
\left(
  \begin{array}{c}
    \psi_1 \\
    \psi_2 \\
  \end{array}
\right)= \left(
  \begin{array}{cc}
    \cos(\chi t) & -i\sin(\chi t)\\
    -i\sin(\chi t) &  \cos(\chi t) \\
  \end{array}
\right)
\left(
  \begin{array}{c}
    \Phi_1 \\
    \Phi_2 \\
  \end{array}
\right),
\label{rabitrans}
\eea
in Eq.~(\ref{lngp1}). This special rotational transformation was first used in Ref.~\cite{rabikev} to convert autonomous Rabi coupled GP system to standard coupled GP system. Following
 this in Ref.~\cite{babujpa} we have applied this to transform non-autonomous Rabi coupled GP system to defocusing GP system. Here the linearly (Rabi) coupled non-autonomous GP system (\ref{lngp1}) is transformed into a
coupled non-autonomous GP system of defocusing type ($\beta(t)>0$) without linear coupling \cite{babujpa}
\bea
i\frac{\partial\Phi_j}{\partial t}=-\frac{1}{2}\frac{\partial^2\Phi_j}{\partial x^2}- \beta(t)\sum_{l=1}^{2}(|\Phi_l|^2)\Phi_j+V_{\mbox{ext}}(x,t)\Phi_j, \quad j=1,2.
\label{ngp}\eea

Next a proper similarity transformation is chosen in order to map the above non-autonomous GP system (\ref{ngp})
to the celebrated Manakov system. The similarity
transformation is given by
\bea
\Phi_j(x,t) = \varepsilon(t)~Q_j\big(X(x,t),T(t)\big)~\mbox{exp}(i \phi(x,t)), \quad j=1,2.
\label{trans-simi}
\eea
Here, the co-ordinates $X(x,t)$ = $\sqrt{2}\sigma_1\left(\beta(t)x-2 \sigma_2 \sigma_1^2\int \beta^2(t)dt\right)$, $T(t)$ = $\sigma_1^2\int \beta^2(t)dt$, $\varepsilon(t)=\sigma_1\sqrt{2\beta(t)}$, and $\phi(x,t)=-\frac{x^2}{2}\left(\frac{\dot{\beta}(t)}{\beta(t)}\right)+2\sigma_1^2\sigma_2\left(\beta(t)x-\sigma_2\sigma_1^2\int \beta^2(t)dt\right)$.
The parameters $\sigma_1$ and $\sigma_2$ are arbitrary real constants.
 Inserting (\ref{trans-simi}) into (\ref{ngp}), we obtain
\bes\bea
i Q_{1,T}+Q_{1,XX}+2(|Q_1|^2+|Q_2|^2)Q_1=0,\\
i Q_{2,T}+Q_{2,XX}+2(|Q_1|^2+|Q_2|^2)Q_2=0,
\eea\label{manakov}\ees
with the condition
\bea
(\ddot{\beta}/\beta)-2(\dot{\beta}/\beta)^2-\Omega^2(t)=0.
\label{riccati}
\eea
The existence of such type of similarity transformation realting non-autonomous system to an integrable autonomous  system was first proposed by Serkin \cite{ser} for the NLS system.
Following this a flurry of activities \cite{non-auto} have been carried out along this direction. However the study of present system (\ref{lngp}) with focusing nonlinearity is still left unexplored.

The similarity transformation between the linearly coupled non-autonomous GP system and the autonomous
 Manakov equation provides us an efficient way to construct
the solutions of the Rabi coupled non-autonomous GP system from the known solutions of the Manakov system \cite{radha}.
The bright one- and two- soliton solutions of the Manakov system (8) are obtained
in Refs.~\cite{radha,kanna2003} are given in the appendix.
In our earlier published work \cite{babujpa}, we have investigated the dynamics of non-autonomous
dark-bright one- and two- soliton solutions under Rabi coupling in the framework of a defocusing coupled NLS system.
In the following section, with the knowledge of the bright one- and two-soliton solutions of the
Manakov system (8) the non-autonomous bright soliton solutions of system (\ref{lngp1})
are obtained and their dynamical properties as well as their stability are discussed in detail.
We would like to emphasize that the bright solitons of the Manakov system admit novel energy sharing
 collisions that have applications in optical computing \cite{computing,babupla}, matter wave
 interferometer \cite{interfero}, and
 partially
coherent solitons \cite{nail, kannaprl} of variable shape. Here we investigate these intriguing collisions in the
 presence of time dependent nonlinearity and Rabi coupling.
\section{Forms of the time-dependent nonlinearity coefficients and their corresponding modulated trap frequencies}
(a)  First we choose a kink-like form of the nonlinearity coefficient,  enabling the transition between two
distinct (constant) values of $\beta$, namely,
\bes\bea
\beta(t) = a_0+\tanh(\rho t+\delta),
\label{kink}
\eea
where $a_0$, $\rho$, and $\delta$ are arbitrary real constants. The associated form of strength of
trap frequency $\Omega^2(t)$ is determined from Eq.~(\ref{riccati})
\bea
\Omega^2(t)=-\frac{2 \rho^2 \mbox{sech}^2(\rho t+\delta)(1+a_0\tanh(\rho t+\delta))}{a_1^2},
\label{kinktr}
\eea\ees
where $a_1 = a_0+\tanh(\rho t+\delta)$.

\begin{figure}[H]
\begin{center}
\vspace{-0.2cm}
\mbox{\hspace{0.2cm}
\subfigure[][]{\hspace{0.1cm}
\includegraphics[height=.27\textheight, angle =0]{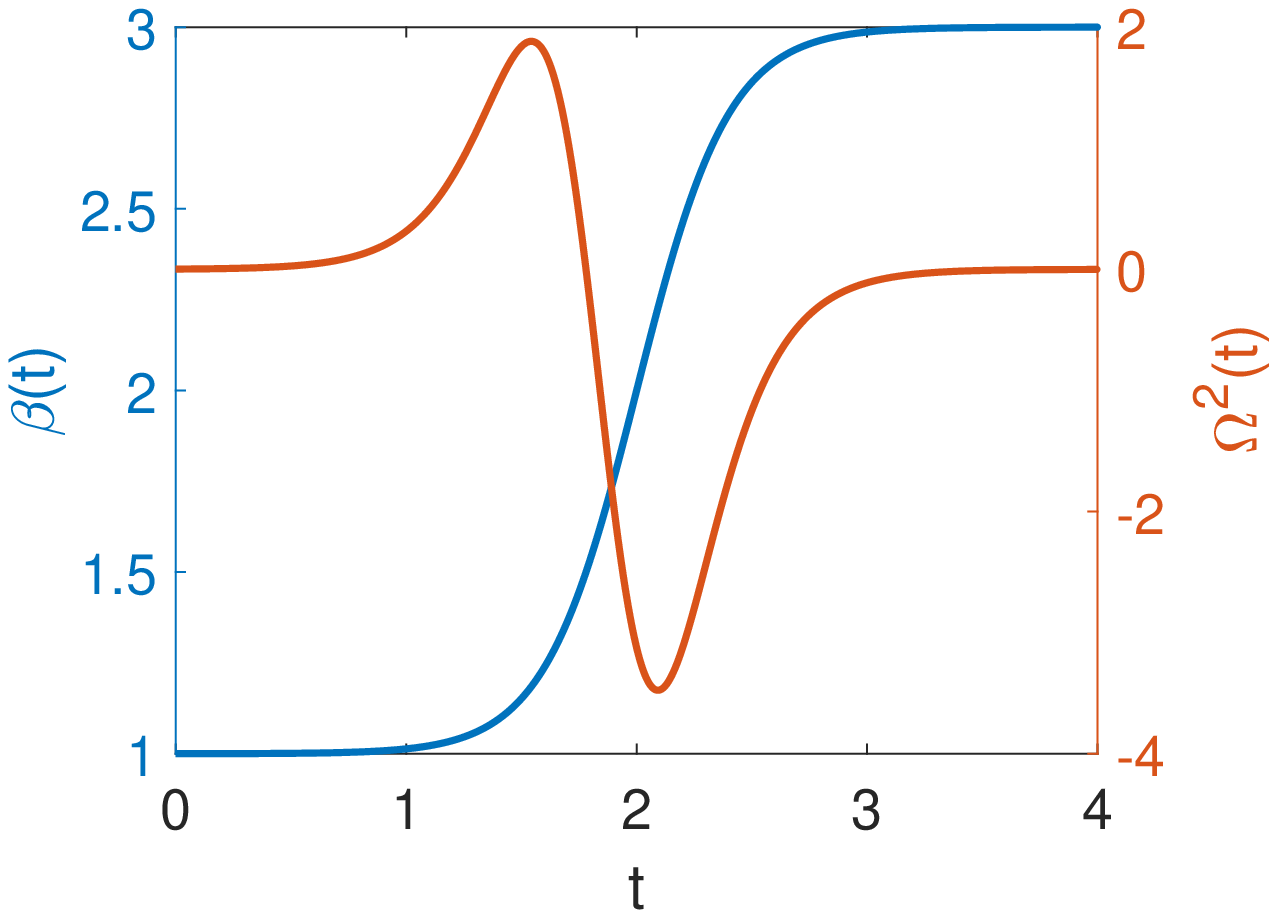}
\label{trap1}
}
\subfigure[][]{\hspace{-1cm}
\includegraphics[height=.26\textheight, angle =0]{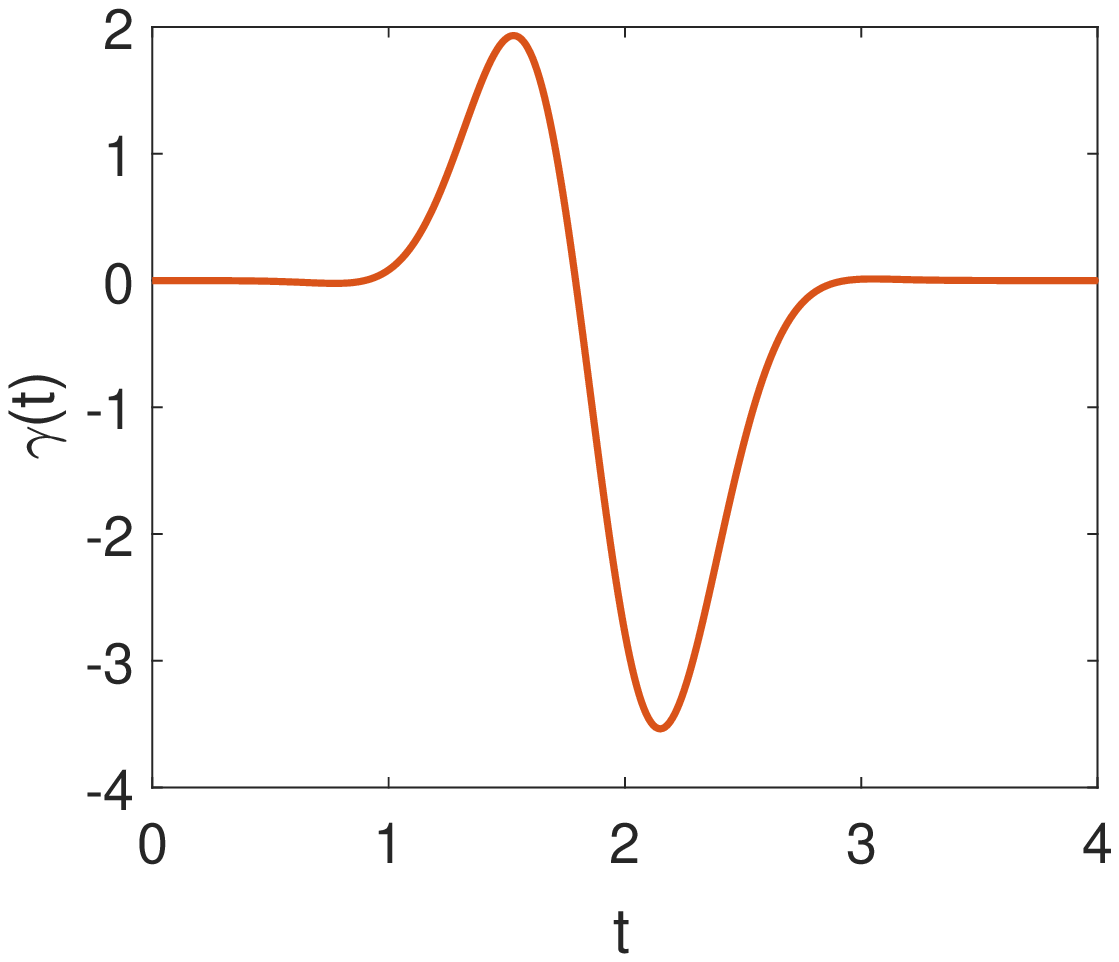}
\label{trap11}
}
}
\end{center}
\caption{Panel (a): Profiles of nonlinearity $\beta(t)$ ($a_0=2$, $\rho=0.5$ and $\delta=-5$) and trap
frequency $\Omega^2(t)$
given by Eqs.~(\ref{kink}) and (\ref{kinktr}), respectively. Panel (b):
Typical plot of $\gamma(t)$ for $w=\sqrt{0.17}$, $c_0=-1.2$, and $c_3=1$.}
\label{trap}
\end{figure}

The graphical structures of $\beta(t)$ and $\Omega^2(t)$ are shown in Fig.~\ref{trap}.
It shows that the temporal modulation of the nonlinearity coefficient $\beta(t)$  (blue solid line)
smoothly varies from one value (lower) to another value (higher) as time $t\rightarrow0$~to~$4$., e.g., in atomic
condensates such variations can be realized by tunning the magnetic field \cite{pollack}.
We depict the nature of the corresponding temporal modulation of the potential $\Omega^2(t)$ in the
same figure as red solid lines. The form of $\Omega^2(t)$ agrees
very well with the Hermite-Gaussian pulse
 $\gamma(t) = [c_0 H_0(\sigma)+c_3 H_3(\sigma)]$exp$(-\sigma^2/2w^2)$, where $\sigma=1.2t-2.25$,
 $w$ is the width of the Gaussian pulse, $c_0$ and $c_3$ are the coefficients of the zeroth order($H_0$)
 and
 third
 order($H_3$)
 Hermite polynomials, respectively. This is clearly shown in Fig.~\ref{trap11}.

(b) We also choose another following form of variable nonlinearity coefficient to examines
\bes\bea
\beta(t) = a_0+\cosh(\rho t+\delta),
\label{coshy}
\eea
where $a_0$, $\delta$ and $\rho$ are, again, real arbitrary constants, while the associated form of the trap
frequency is found from Eq.~(\ref{riccati}) as
\bea
\Omega^2(t)=\frac{\rho^2 [3+2a_0\cosh(\rho t+\delta)-\cosh(2(\rho t+\delta))]}{2a_1^2},
\label{cosphy1}
\eea\ees
where $a_1 = a_0+\cosh(\rho t+\delta)$.

\begin{figure}[H]
\begin{center}
\vspace{-0.2cm}
\mbox{\hspace{0.2cm}
\subfigure[][]{\hspace{0.1cm}
\includegraphics[height=.27\textheight, angle =0]{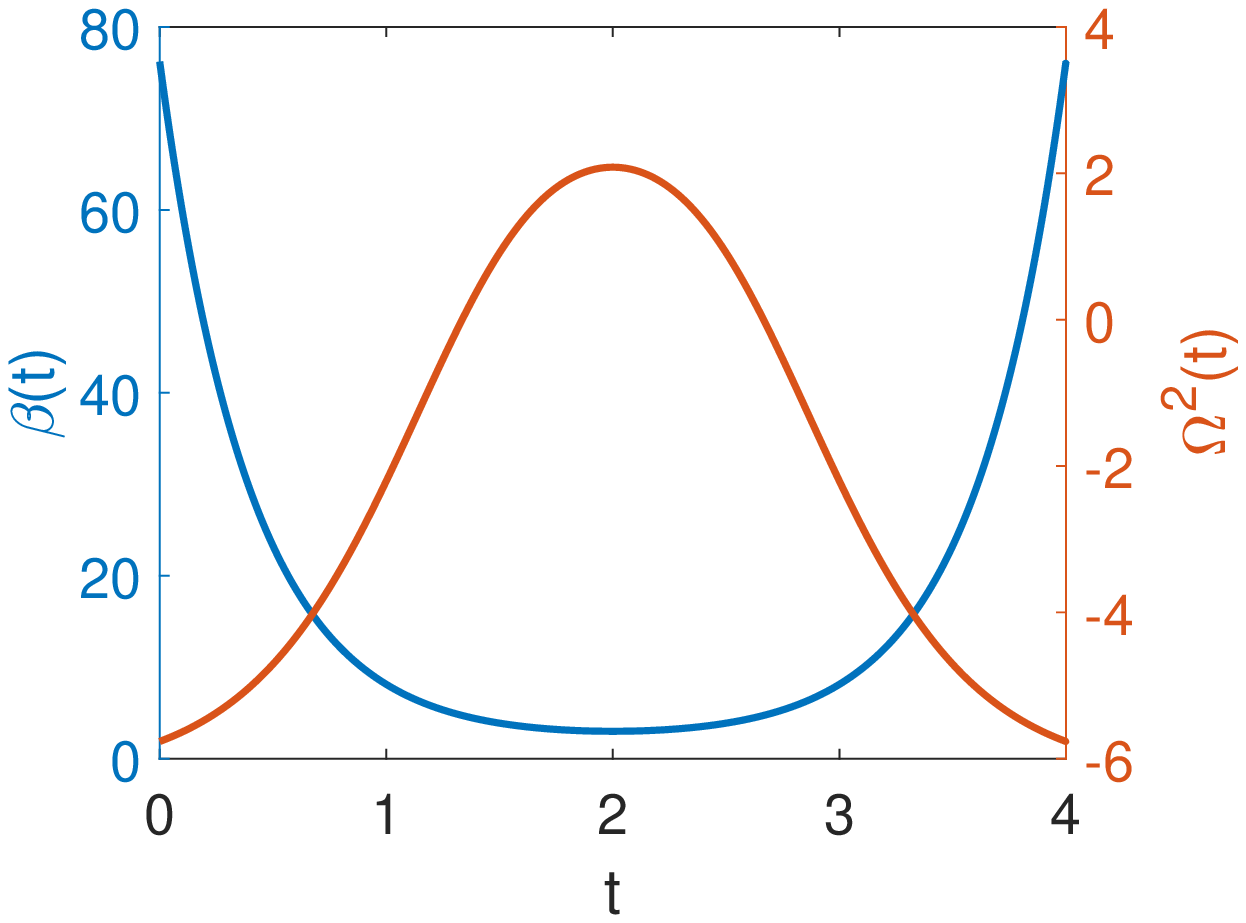}
\label{trap2}
}
\subfigure[][]{\hspace{-1cm}
\includegraphics[height=.26\textheight, angle =0]{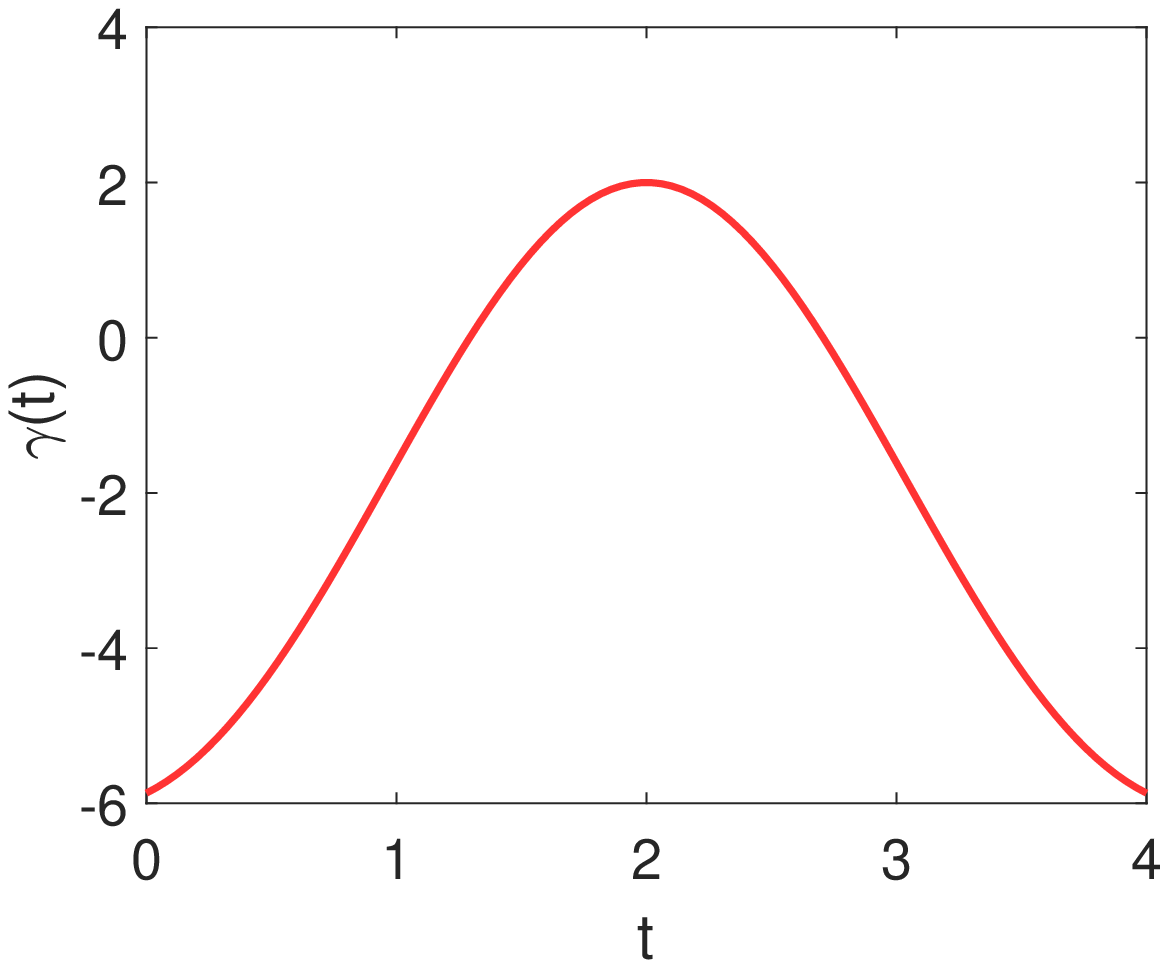}
\label{trap3}
}
}
\end{center}
\caption{Panel (a): Profile of nonlinearity $\beta(t)$($a_0=2$, $\rho=2.5$, and $\delta=-5$) and
trap frequency $\Omega^2(t)$ given by Eqs.~(\ref{coshy}) and (\ref{cosphy1}).
Panel (b): Typical plot of $\gamma(t)$ for $w=\sqrt{2.3}$ and $a=-1$.}
\label{trap_1}
\end{figure}

The graphical structures of $\beta(t)$ and $\Omega^2(t)$ are sketched in Fig.~\ref{trap_1}. The nature
of time-dependent function $\beta(t)$ (blue solid line) admits a flat bottom parabolic profile (at $t=0$ and $4$ it
reaches its maximum value, while $t=2$ approaches its minimum value). Indeed it changes its value from negative to positive which can be very well achieved through FR mechanism.
Next, the temporal modulation of the trap potential $\Omega^2(t)$ is found to be localized pulse,
which can be experimentally realized. It is interesting to note that the nature of the above
function $\Omega^2(t)$ well agrees with the function $\gamma(t)=aH_2(\sigma)$exp$(-\sigma^2/2w^2)$,
where $\sigma=t-2$, $a$ and $w$ are the amplitude and width of the Gaussian pulse/beam,
$H_2(\sigma)$ is the second order Hermite polynomial (see Fig.~\ref{trap3}).
We believe that this resemblance of $\Omega^2(t)$ with Hermite-Gaussian (HG) pulse (second order)
pointed out here, will pave way to realize non-autonomous solitons experimentally in multiple species
condensates as these HG pulses can be formed by suitable laser sources.
\section{Explicit soliton solutions of non-autonomous coupled GP system }
Here we write down the one- and two-soliton solutions of system (\ref{lngp1}) for constant as well as
time-dependent nonlinearity coefficients. This will enable us to compare the dynamics of the bright
solitons in the presence and absence of the time dependence of the nonlinearity coefficient $\beta$.
\subsection{\bf Bright one- soliton solution}
  (i) The bright one-soliton solution of system (\ref{lngp1}) for constant nonlinearity coefficient $\beta=1$
  and in the absence of an external potential $V_{\mbox{ext}}(x,t)$ is given by \cite{kanna2003}
  \bes\bea
\psi_1(x,t)& =& \left(A_1\cos(\chi t)Q_1-iA_2\sin(\chi t)Q_2\right),\\
\psi_2(x,t)& =& \left(-iA_1\sin(\chi t)Q_1+A_2\cos(\chi t)Q_2\right),
\label{soliton}
\eea\ees
where the functions $Q_1$ and $Q_2$ are given in Eq.~(\ref{b-one}) in the appendix.
Here the variables $\omega$ and $\eta_{1I}$ are redefined as
$\omega$ = $x-k_{1I}t$ and $\eta_{1I}$ = $k_{1I} x+(k_{1R}^2-k_{1I}^2)(t/2)$.

For constant nonlinearity coefficient (i.e., $\beta(t)$ = const.) and in the absence of external potential
($V_{\mbox{ext}}=0$, corresponding
to homogeneous condensate), the soliton profiles in both the components $\psi_1(x,t)$ and $\psi_2(x,t)$
exhibit breathing (or) oscillating behavior due to the Rabi coupling parameter $\chi$.
\begin{figure}[H]
\begin{center}
\vspace{-0.2cm}
\mbox{\hspace{-0.2cm}
\subfigure[][]{\hspace{-0.2cm}
\includegraphics[height=.25\textheight, angle =0]{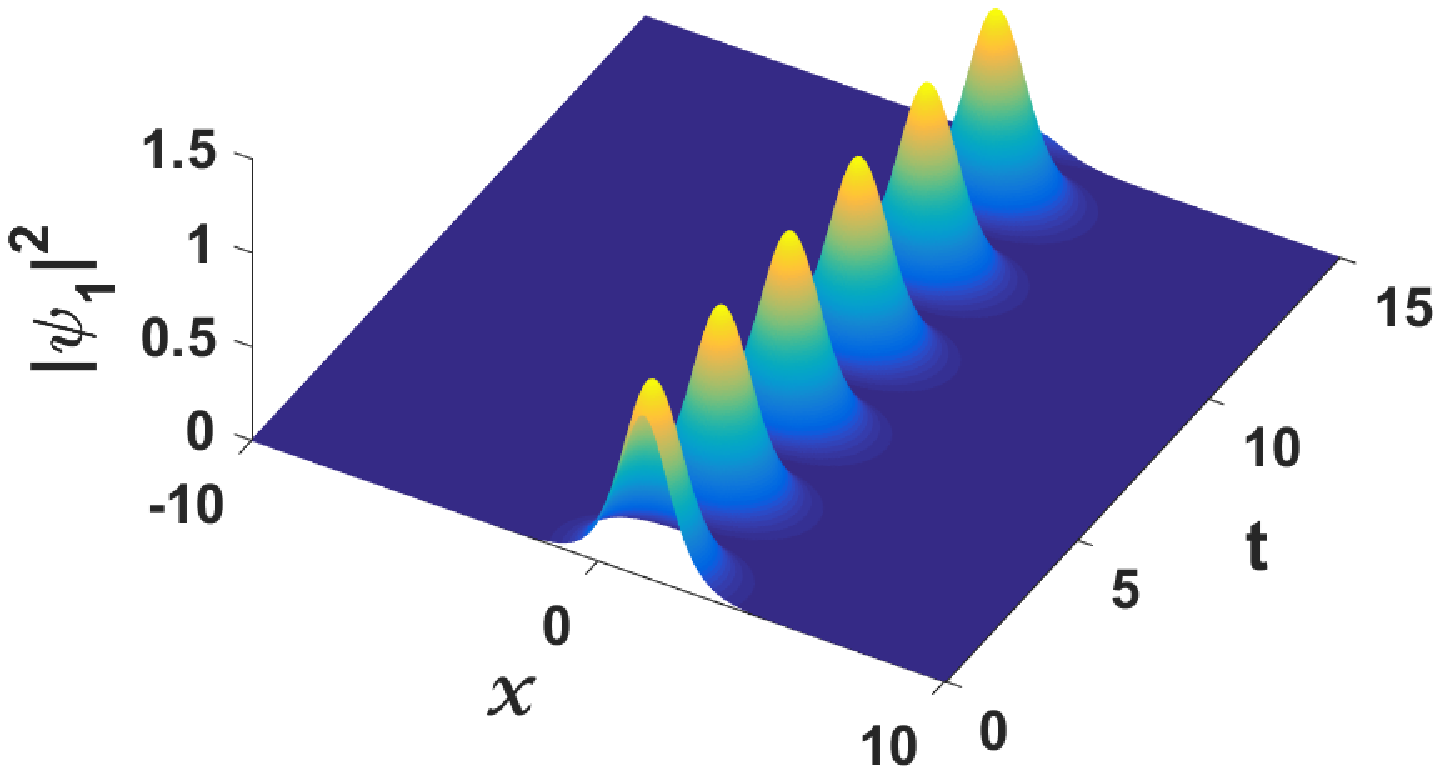}
\label{fig1a}
}
\subfigure[][]{\hspace{-0.2cm}
\includegraphics[height=.25\textheight, angle =0]{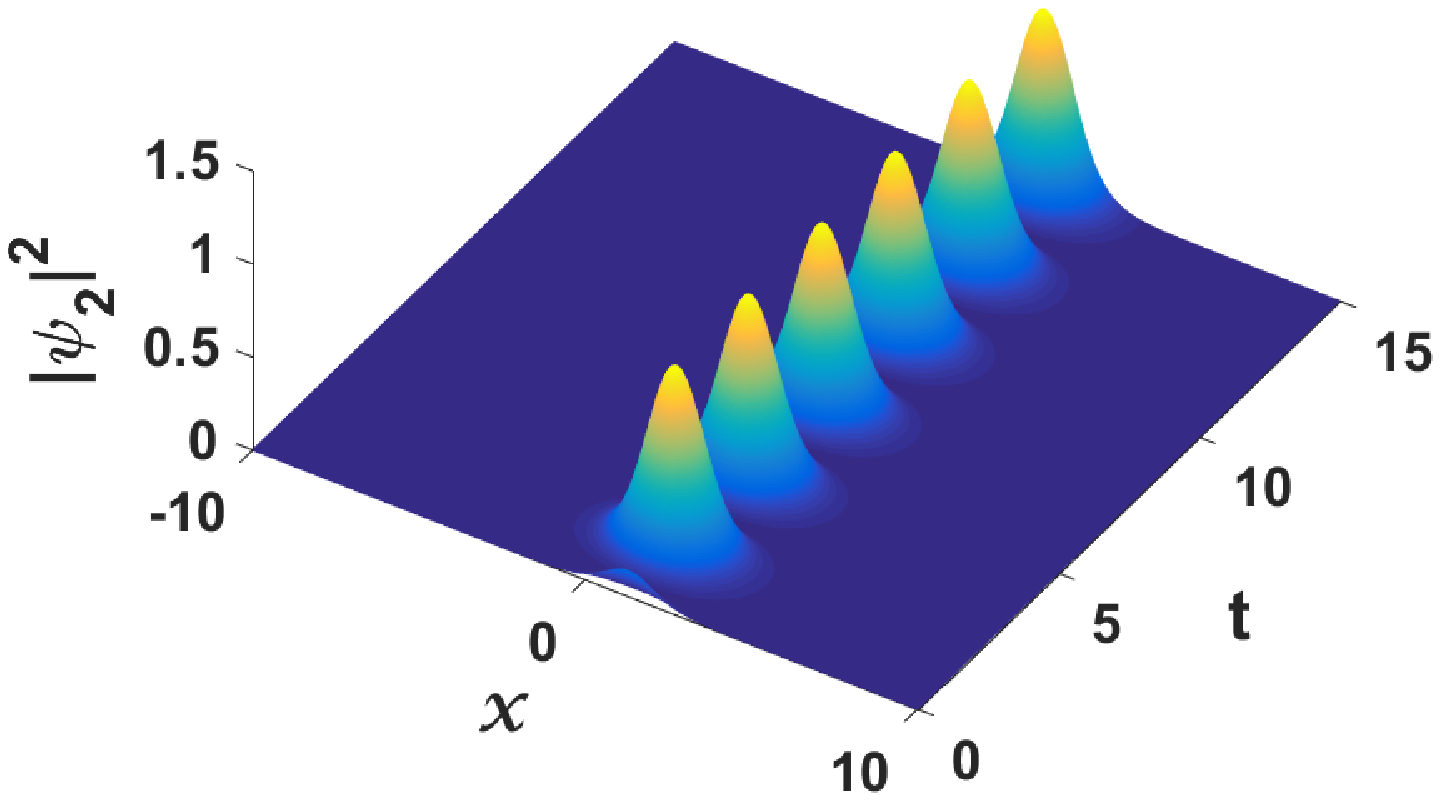}
\label{fig1b}
}
}
\subfigure[][]{\hspace{-0.2cm}
\includegraphics[height=.26\textheight, angle =0]{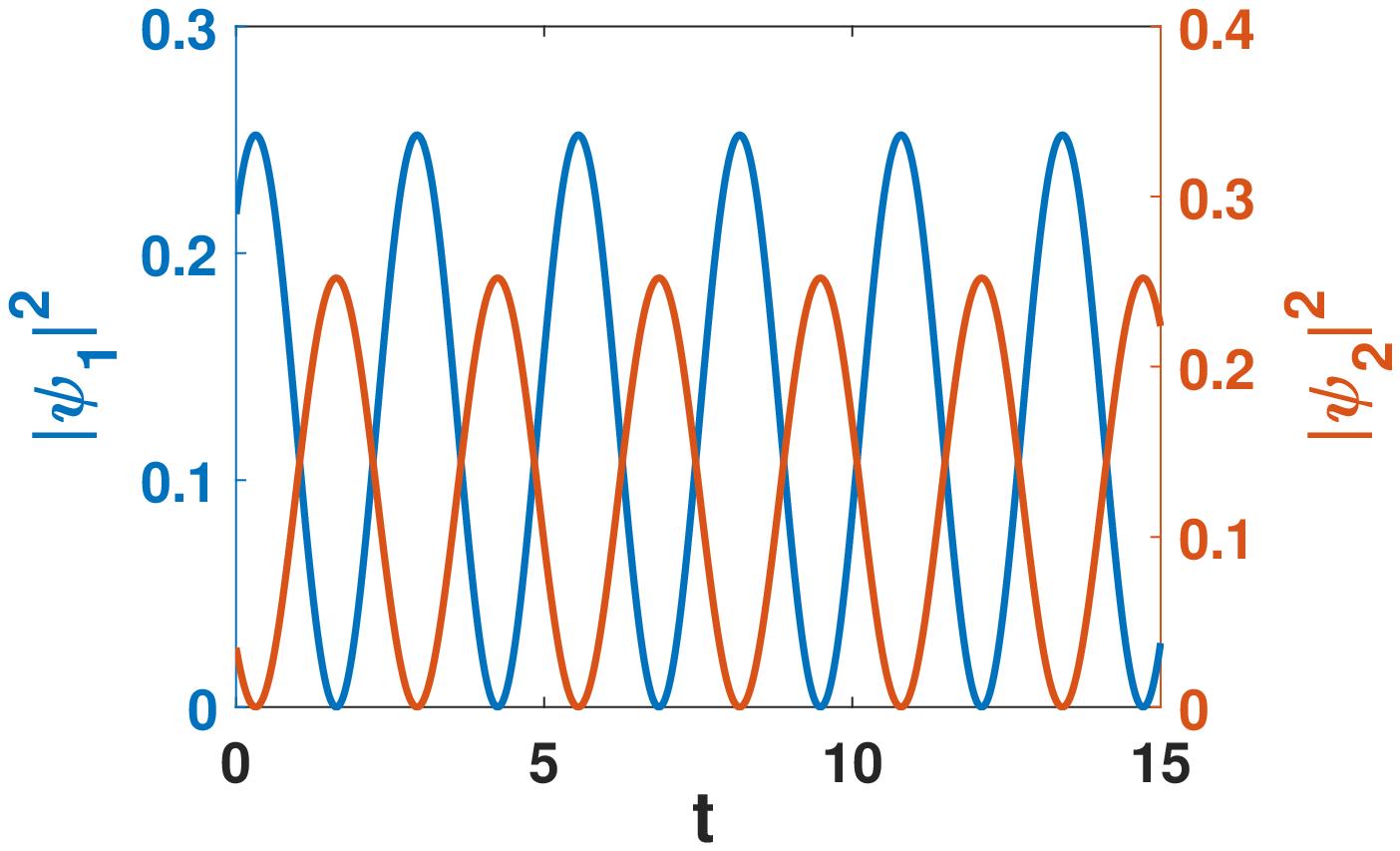}
\label{fig1b}
}
\end{center}
\caption{Oscillating soliton for constant nonlinearity coefficient $\beta(t)$ and in the absence of external potential.
Panels (a) and (b) show contour plots of density profiles of $|\psi_1|^2$ and $|\psi_2|^2$. Panel (c) shows
density profiles of $|\psi_1|^2$ and $|\psi_2|^2$. Here $k_{1R}=1$, $k_{1I}=0$, $\alpha_1^{(1)}=0.5$, $\beta=1$,
$\alpha_1^{(2)}=0.2i$, and $\chi=1.2$.}
\label{fig1}
\end{figure}
 Figs.~\ref{fig1a}-\ref{fig1b}
show a train of identical breathing solitons/breathers. To facilitate the understanding of such oscillations we
explicitly present the expressions for the condensate densities
$|\psi_1|^2 = k_{1R}^2\mbox{sech}^2(k_{1R}\omega+R/2)(|A_1|^2\cos^2(\chi t)+|A_2|^2\sin^2(\chi t))+i k_{1R}^2\cos(\chi t)\sin(\chi t)\mbox{sech}^2(k_{1R}\omega+R/2)(A_1A_2^*-A_1^*A_2)$ and $|\psi_2|^2 = k_{1R}^2\mbox{sech}^2(k_{1R}\omega+R/2)(|A_2|^2\cos^2(\chi t)+|A_1|^2\sin^2(\chi t))+i k_{1R}^2\cos(\chi t)\sin(\chi t)\mbox{sech}^2(k_{1R}\omega+R/2)(A_1^*A_2-A_1A_2^*)$.
 From this expression we find that the oscillations originate from the trigonometric
 functions ($cosine$ and $sine$) appearing in the cross (interference) terms. The matter wave
 oscillation between the components becomes larger for large values of the Rabi coupling $\chi$ as expected.\\

\begin{figure}[H]
\begin{center}
\vspace{-0.2cm}
\mbox{\hspace{0.6cm}
\subfigure[][]{\hspace{-1.0cm}
\includegraphics[height=.25\textheight, angle =0]{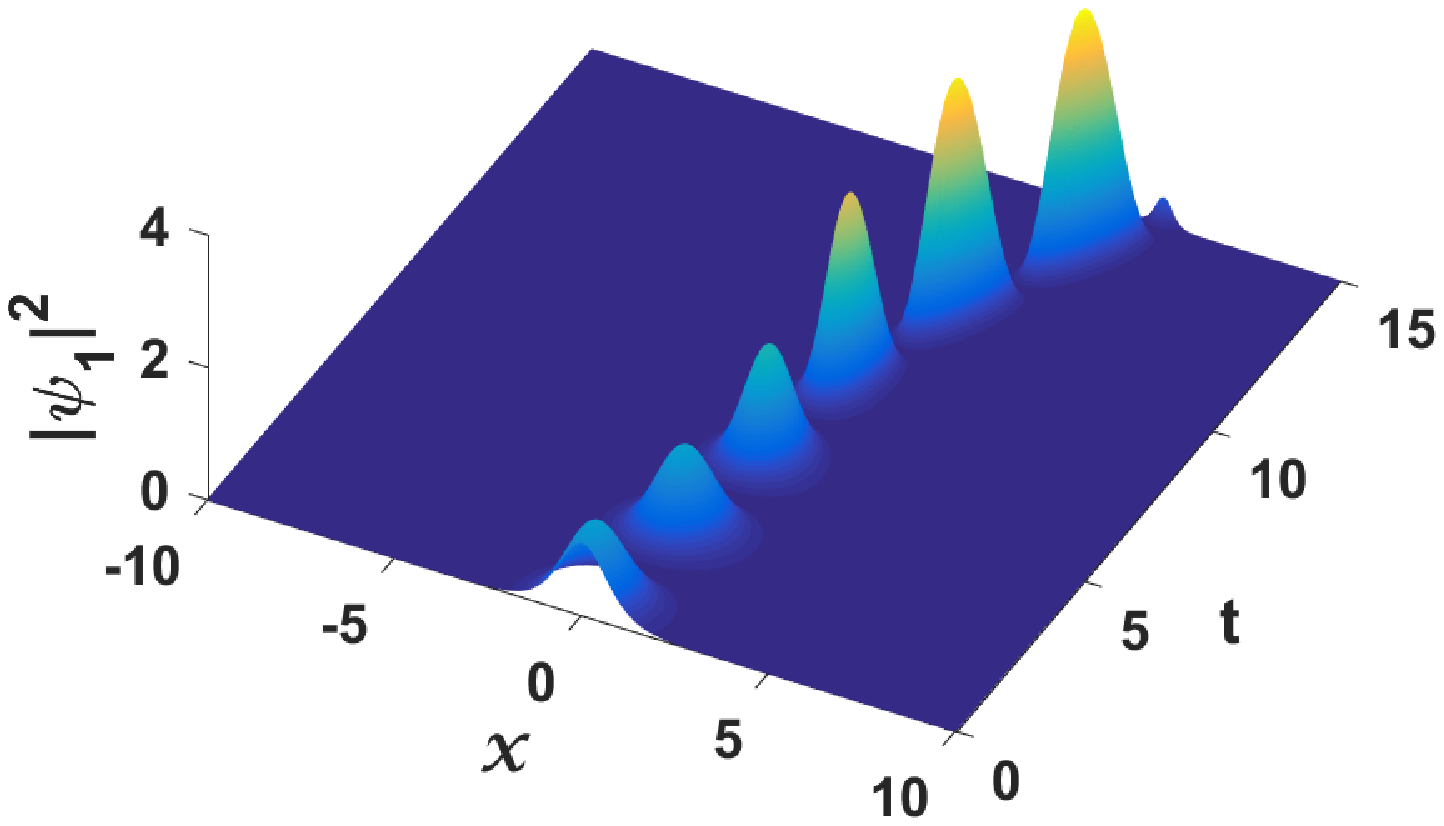}
\label{fig1c}
}
\subfigure[][]{\hspace{-0.2cm}
\includegraphics[height=.25\textheight, angle =0]{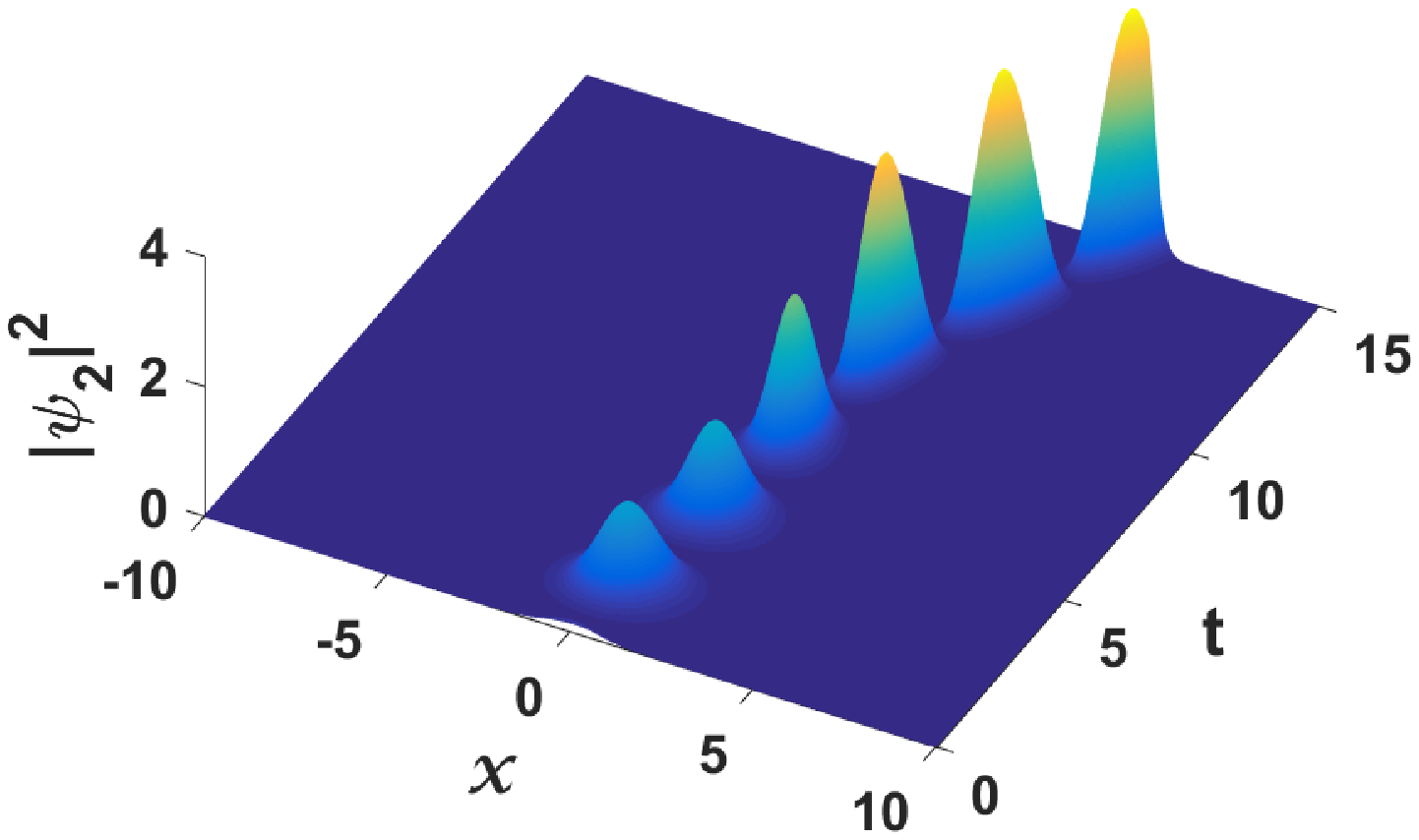}
\label{fig1d}
}
}
\mbox{\hspace{0.5cm}
\subfigure[][]{\hspace{-1.0cm}
\includegraphics[height=.25\textheight, angle =0]{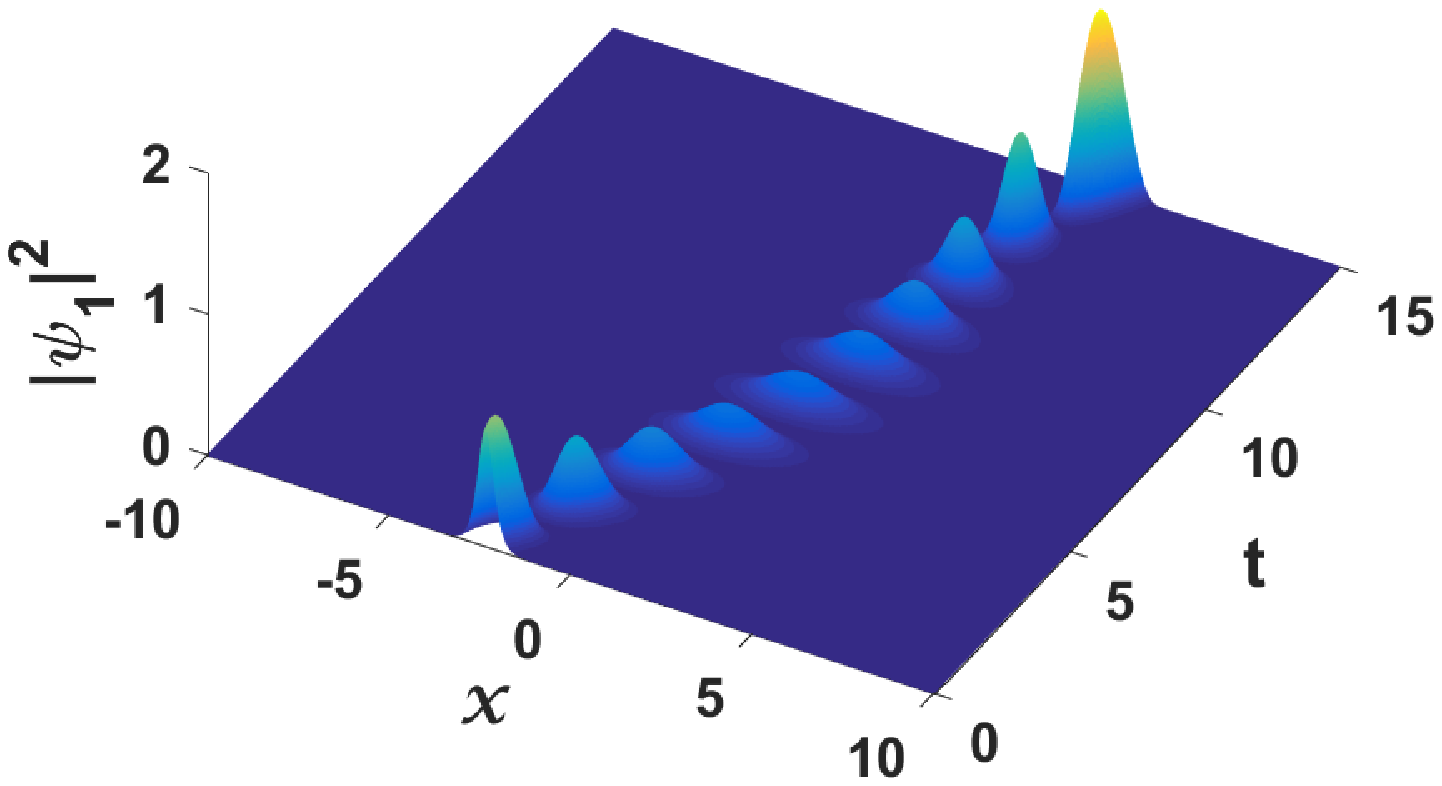}
\label{fig1e}
}
\subfigure[][]{\hspace{-0.2cm}
\includegraphics[height=.25\textheight, angle =0]{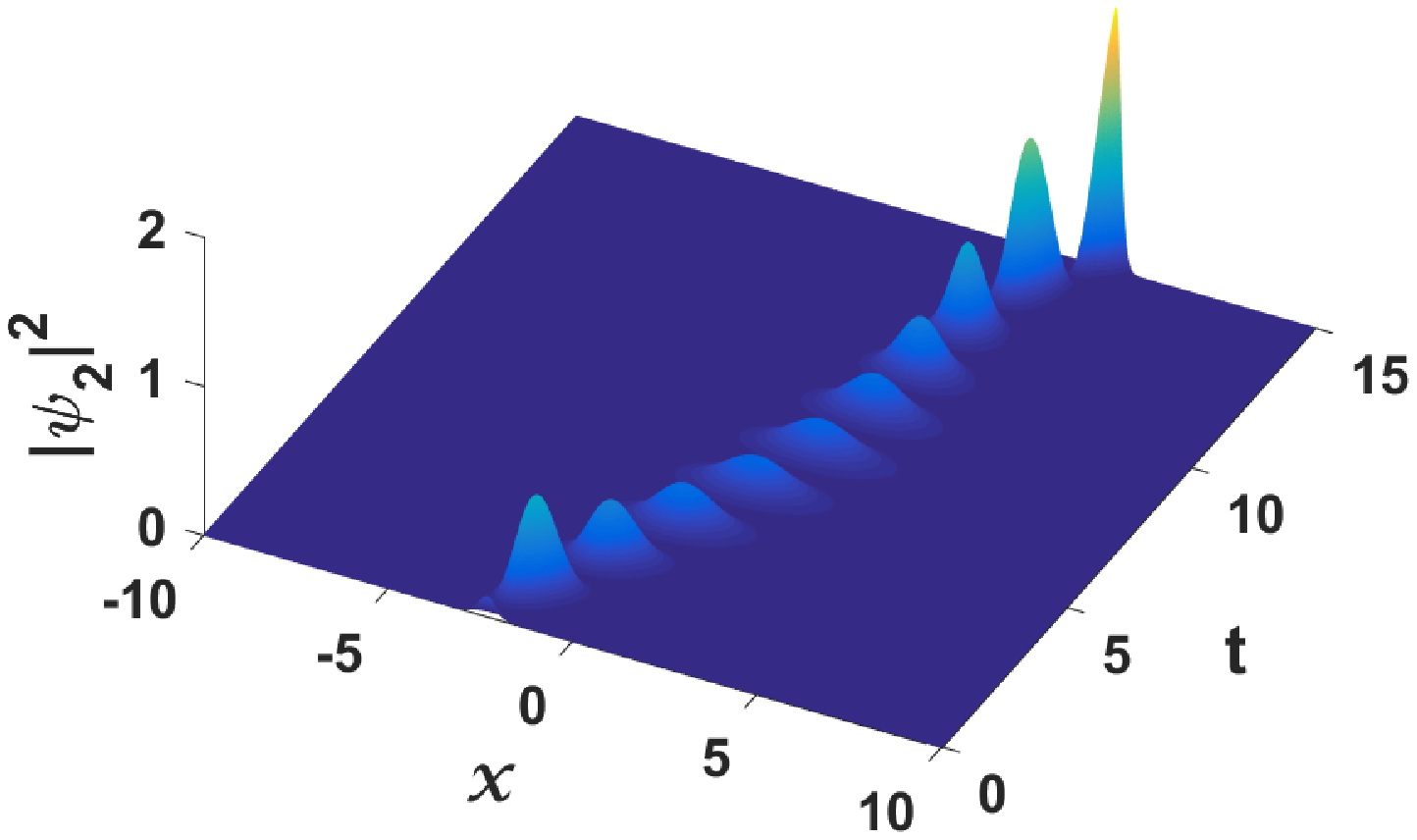}
\label{fig1f}
}
}
\end{center}
\caption{Time-dependent nonlinearity coefficient $\beta(t)$ and presence of external
potential (top and bottom row panels). Panels (a) and (b) correspond to values of
 parameters of $\beta=a_0+\tanh(\rho t+\delta)$, $k_{1I}=0.2$, $\rho=0.4$,
 $\sigma_1=0.8$, $\delta=-3$, $\chi=1.2$, $a_0=2$, and $\sigma_2=0$,
 whereas (c) and (d) correspond to values of parameters $\beta=a_0+\cosh(\rho t+\delta)$,
 $k_{1I}=0.5$, $\rho=0.45$, $\delta=-3$, $\sigma_1=0.2$, $\chi=2$, $a_0=2$, and $\sigma_2=0$.
 All other values of soliton parameters are same as given in Figs.~\ref{fig1a}-\ref{fig1b}.}
\label{fig2}
\end{figure}

(ii) The non-autonomous bright one-soliton solution of system (\ref{lngp1}) for time-dependent nonlinearity
coefficient $\beta(t)$ and in the presence of external potential $V_{\mbox{ext}}(x,t)$ obtained by using the
transformations mentioned in the previous section and the solution given in the appendix, is given below
 \bes\bea
\hspace{-1cm}\psi_1(x,t)& =& \left[A_1\cos(\chi t)-iA_2\sin(\chi t)\right]\varepsilon(t)k_{1R}~\mbox{sech}(k_{1R}\tilde{\omega}+R/2)e^{i(\tilde{\eta}_{1I}+\phi(x,t))},\\
\hspace{-1cm}\psi_2(x,t)& = &\left[-iA_1\sin(\chi t)+A_2\cos(\chi t)\right]\varepsilon(t)k_{1R}~\mbox{sech}(k_{1R}\tilde{\omega}+R/2)e^{i(\tilde{\eta}_{1I}+\phi(x,t))},
\eea
where
\bea
\hspace{-1cm}\tilde{\omega}&=&X-2k_{1I}T\\
\hspace{-1cm}\tilde{\omega}&=&\sqrt{2}\sigma_1\beta(t) x-2\sqrt{2}\sigma_1^3\sigma_2 \int \beta^2(t)dt-2k_{1I}
\sigma_1^2\int \beta^2(t)dt,\\
\hspace{-1cm} \tilde{\eta}_{1I}&=&\sqrt{2}\sigma_1k_{1I}\beta(t) x-2\sqrt{2}k_{1I}\sigma_1^3
\sigma_2 \int \beta^2(t)dt+(k_{1R}^2-k_{1I}^2)\sigma_1^2\int \beta^2(t)dt,
\eea
and
\bea
\phi(x,t)=\left(-\frac{\dot{\beta}(t)}
{2\beta(t)}x^2+2\sigma_1^2\sigma_2(\beta(t)x-\sigma_2\sigma_1^2\int \beta^2(t)dt)\right),
\eea\label{nonsoliton}\ees
where the overdot denotes $d/dt$. All other parameters are given below Eq.~(\ref{b-one}) in the appendix.
The mesh plot of evolution of a single oscillating non-autonomous bright soliton is shown in Fig.~\ref{fig2}.

 In this case, we consider the nonlinearity coefficient to be time-dependent and include the external
 potential for studying the soliton dynamics. For this purpose, we choose two types of time-dependent
 nonlinearity coefficients in terms of hyperbolic functions discussed in the previous section.
 Figs.~\ref{fig1c}-\ref{fig1d} show the matter wave soliton compression for the choice
 $\beta(t)=a_0+\tanh(\rho t+\delta)$.
At time $t\rightarrow0$, the amplitude and width of the soliton is lower and wider.
As time goes on, the amplitude is gradually increased and the pulse width is narrowed down (see time $t\rightarrow15$).
 Note that the velocity of the oscillating soliton is strongly affected by the nature of the time-dependent nonlinearity.
Next, we consider the form of time-dependent nonlinearity coefficient as $\beta(t)=a_0+\cosh(\rho t+\delta)$.
In this case the soliton is oscillating and the amplitude is higher at $t=0$ and at $t=15$. But in between the
amplitude is lower. Note that the central position of the soliton also oscillates periodically.
We view this as an oscillating soliton cradle. This is clearly sketched in Figs.~\ref{fig1e}-\ref{fig1f}.

\subsection{\bf Bright two-soliton solution and soliton collision}

\begin{figure}[H]
\begin{center}
\vspace{-0.5cm}
\mbox{\hspace{0.4cm}
\subfigure[][]{\hspace{-0.2cm}
\includegraphics[height=.25\textheight, angle =0]{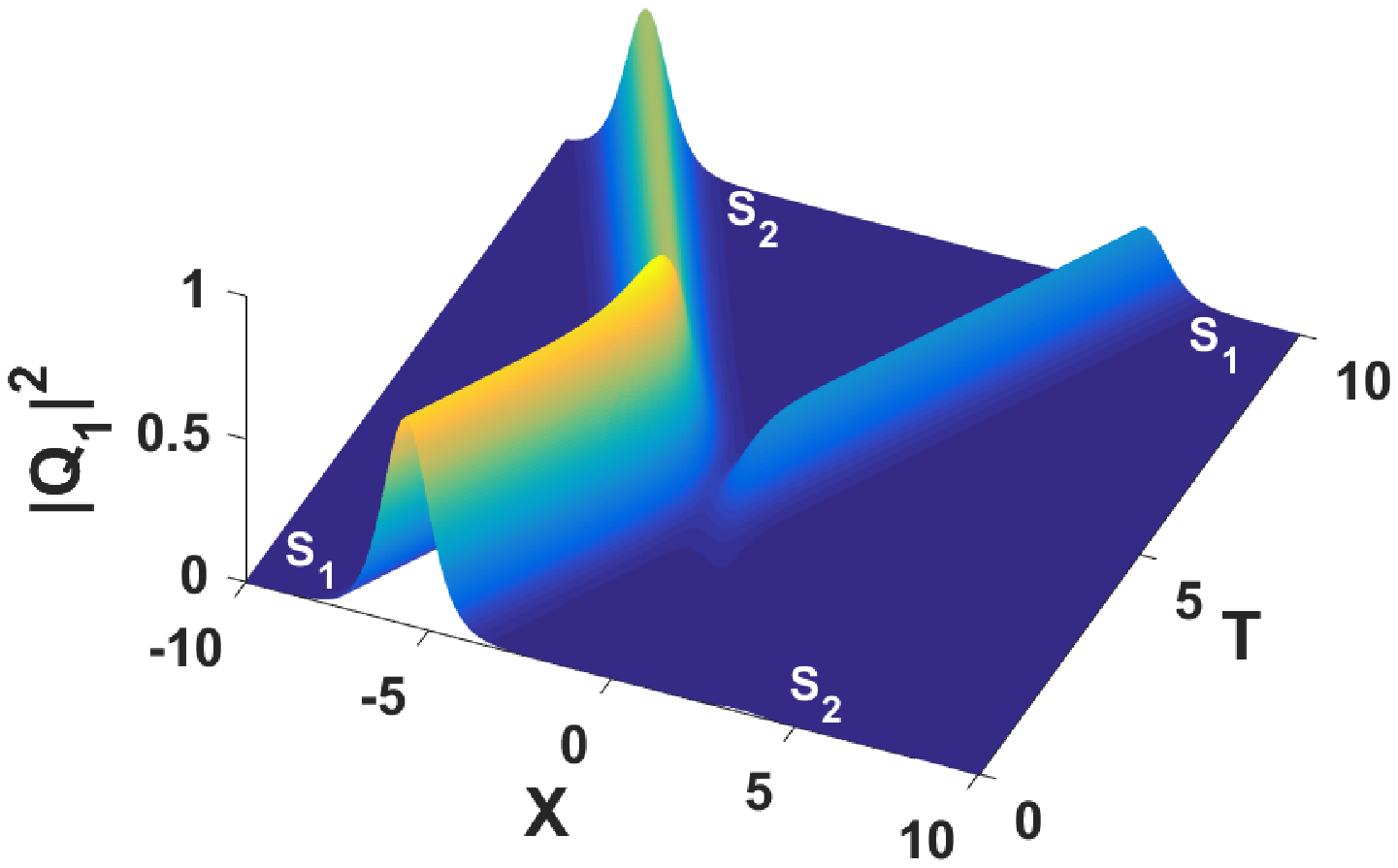}
\label{fig2a}
}
\subfigure[][]{\hspace{-0.2cm}
\includegraphics[height=.25\textheight, angle =0]{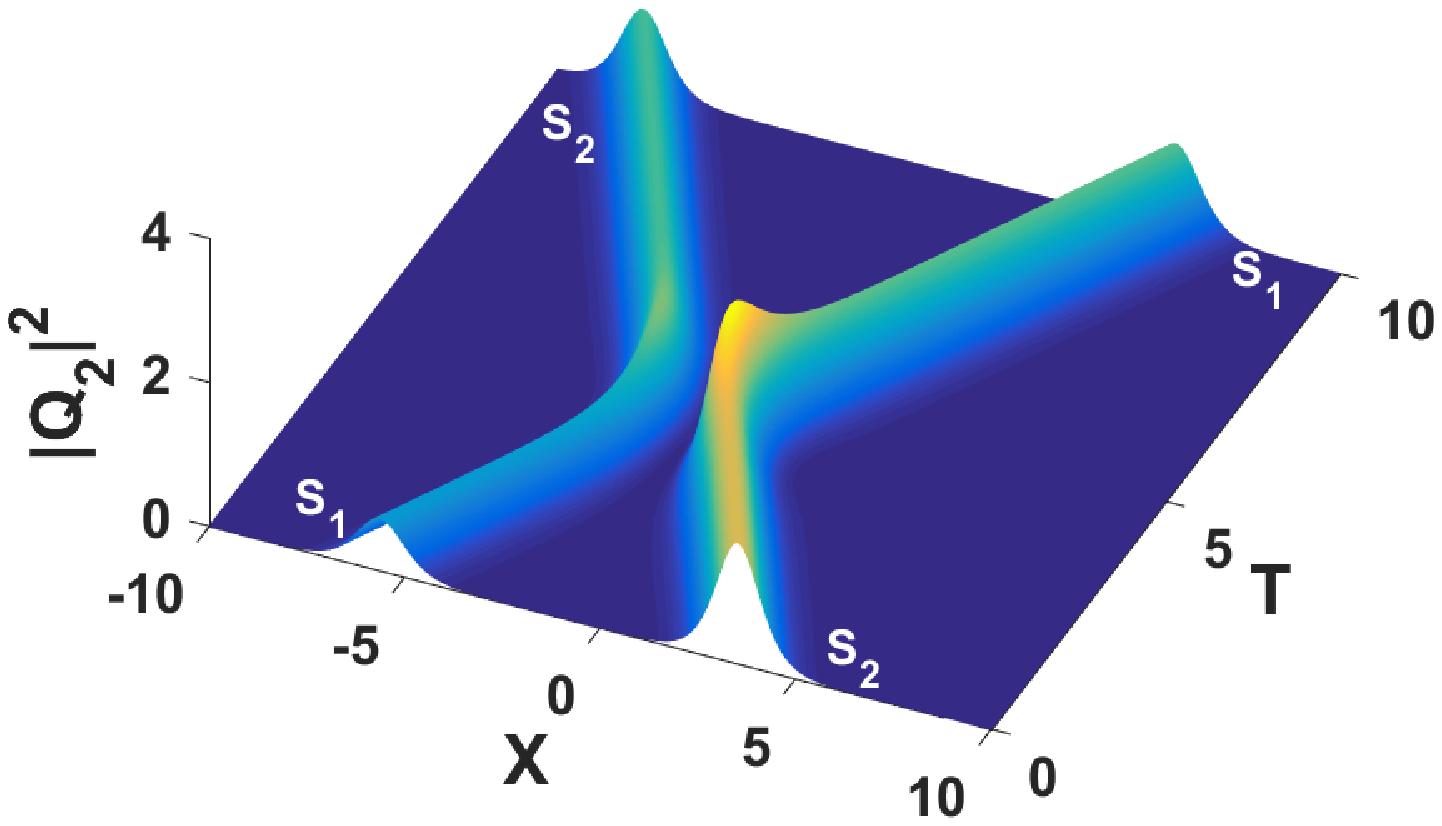}
\label{fig2b}
}
}
\end{center}
\caption{Panels (a)-(b): Shape changing/energy sharing collision of Manakov system (8)
with parameters $\alpha_1^{(1)}=\alpha_1^{(2)}=\alpha_2^{(1)}=0.002$, $\alpha_2^{(2)}=0.01+0.025i$,
$k_1=-1.2+0.5i$, and $k_2=1.3-0.5i$.}
\label{fig3}
\end{figure}

{\bf (i) Brief revisit of collision in the Manakov system}:

The soliton solution of Manakov system (8) is described by the two-soliton solution given in appendix
(see Eq.~(\ref{two-sol})).
 The nature of the two soliton collision is shown in Figs.~\ref{fig2a}-\ref{fig2b}.
 It shows the shape changing (energy sharing) collision of solitons of system (8) \cite{kanna2003}.
 Here, $S_1$ and $S_2$ denote the first and second soliton, respectively. In the $Q_1(X,T)$ component the condensate
 density of soliton $S_1$ gets suppressed after collision while there is an enhancement in that of second
 soliton $S_2$. The reverse scenario takes place in the $Q_2(X,T)$ component.
 See Ref.~\cite{pramana} for a review of this energy sharing collision.

{\bf (ii) Breather production in soliton collision}

\begin{figure}[H]
\begin{center}
\vspace{-1cm}
\mbox{\hspace{0.5cm}
\subfigure[][]{\hspace{-1.0cm}
\includegraphics[height=.25\textheight, angle =0]{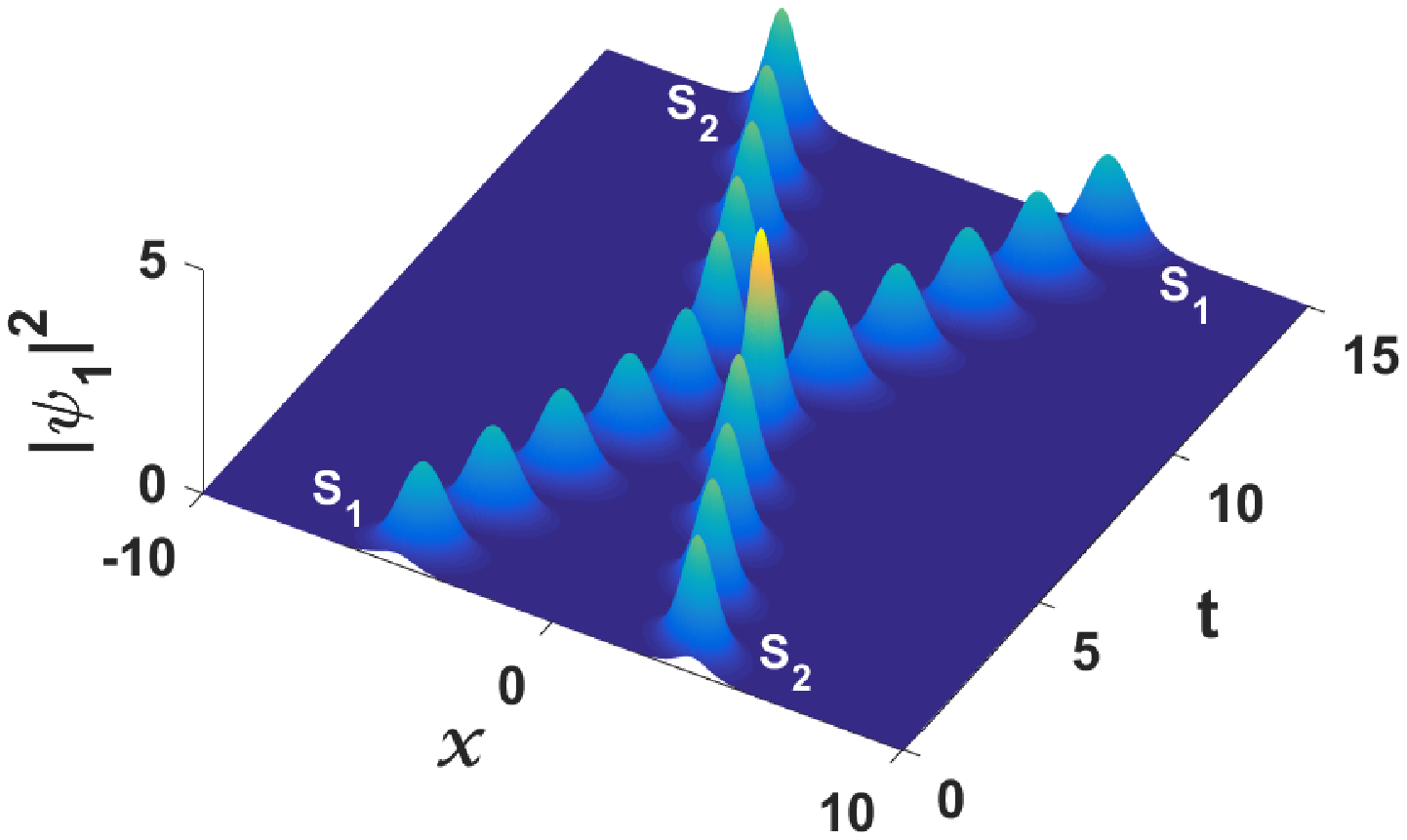}
\label{fig2c}
}
\subfigure[][]{\hspace{-0.2cm}
\includegraphics[height=.25\textheight, angle =0]{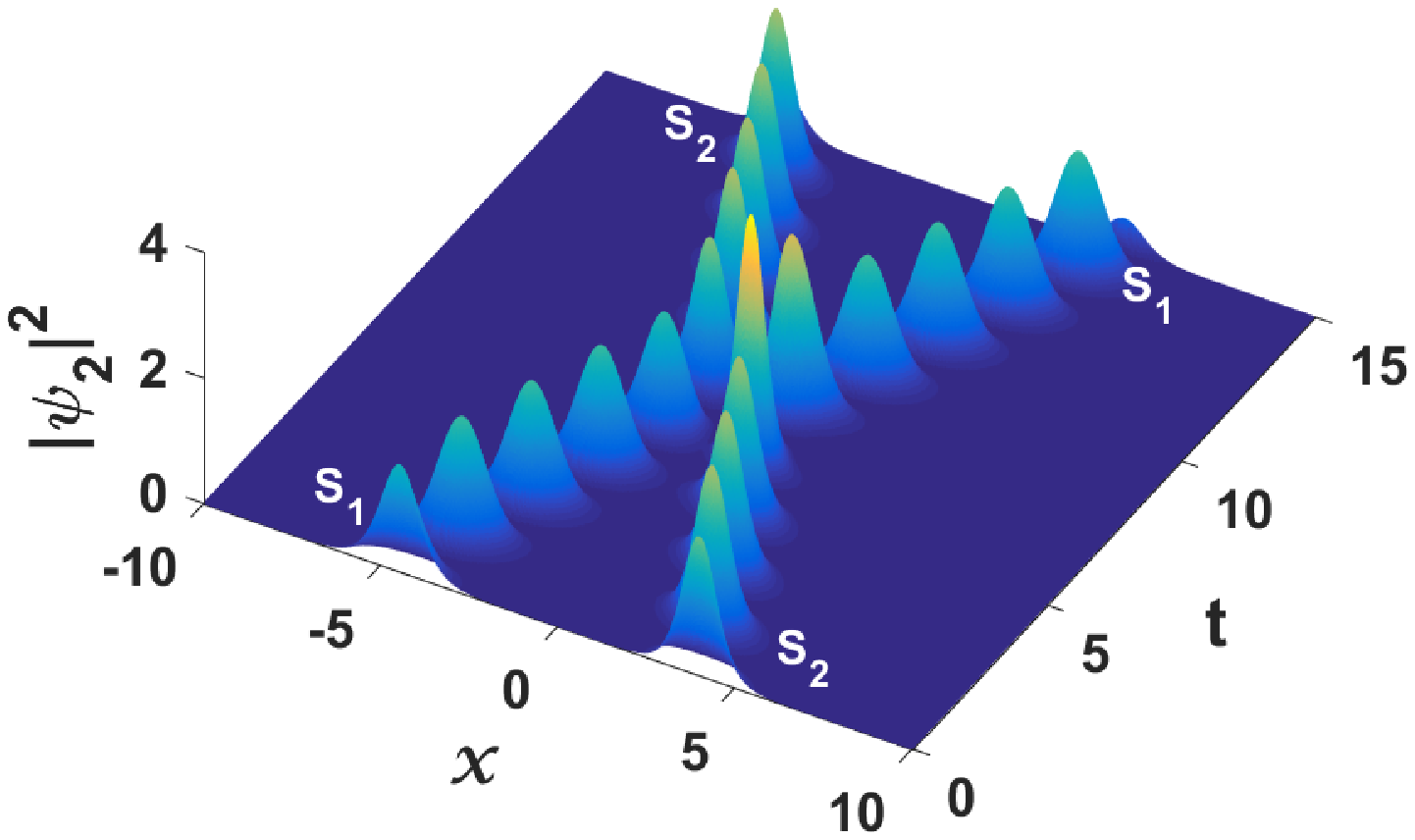}
\label{fig2d}
}
}
\mbox{\hspace{0.5cm}
\subfigure[][]{\hspace{-1.0cm}
\includegraphics[height=.25\textheight, angle =0]{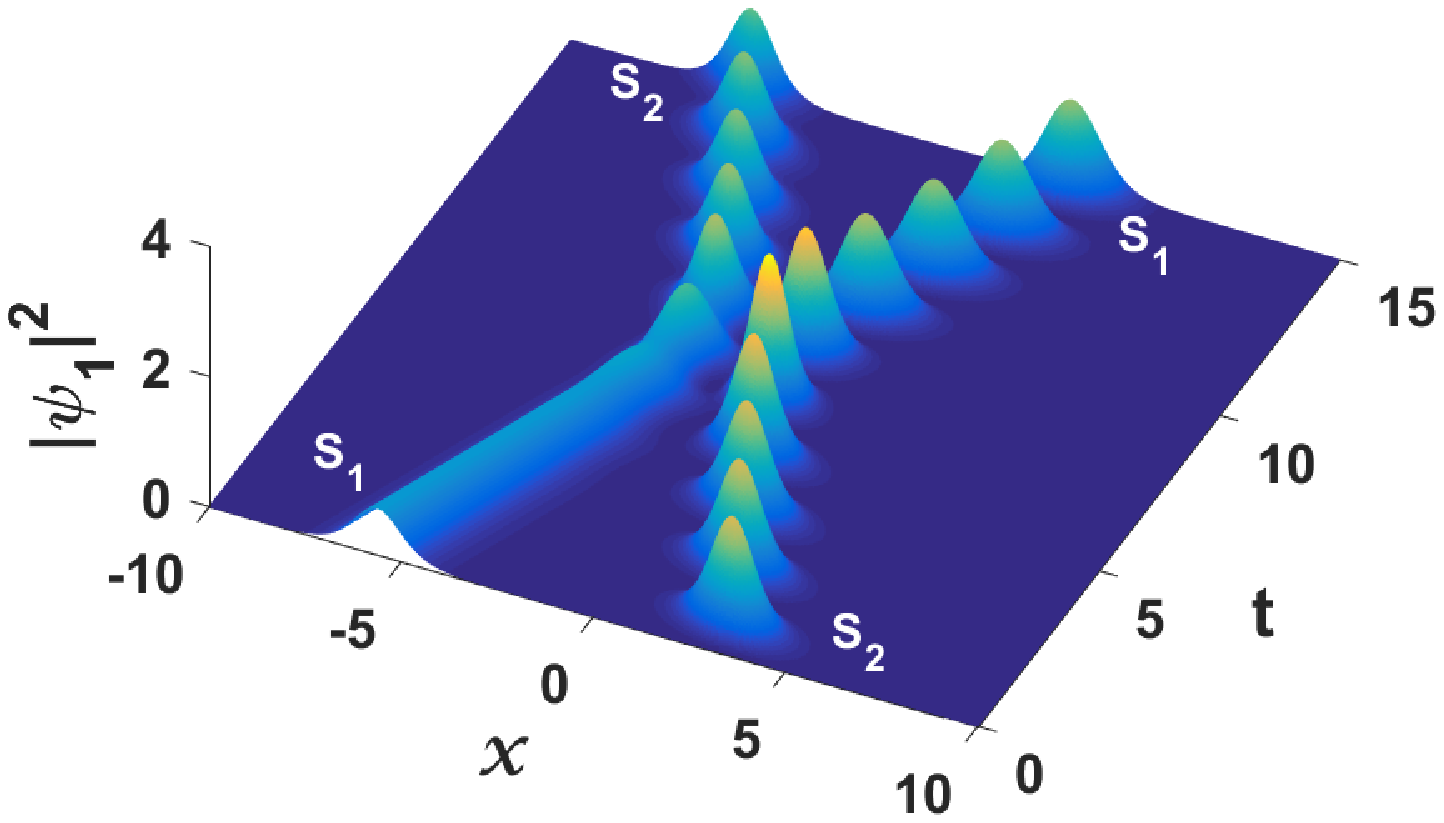}
\label{fig2e}
}
\subfigure[][]{\hspace{-0.2cm}
\includegraphics[height=.25\textheight, angle =0]{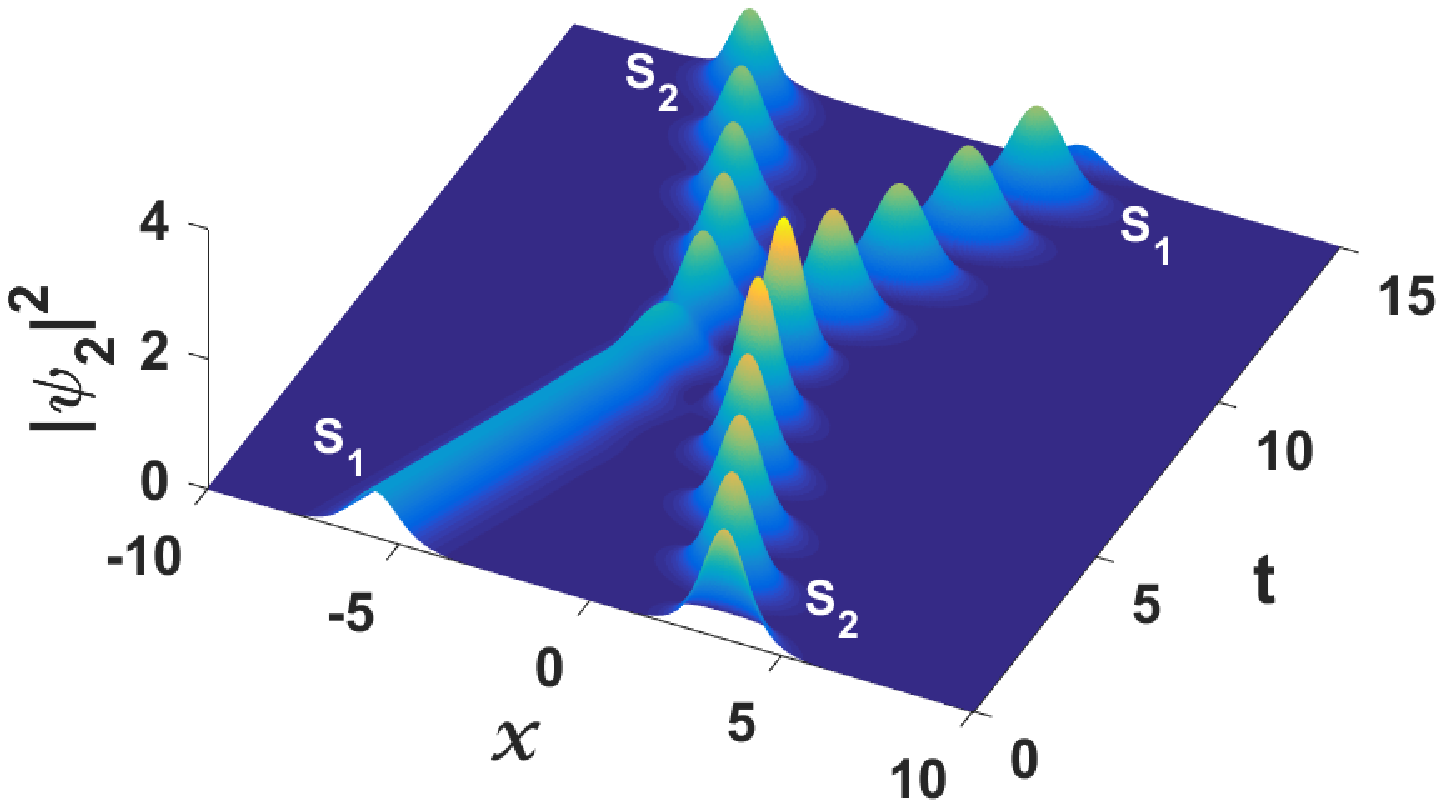}
\label{fig2f}
}
}
\subfigure[][]{\hspace{-0.8cm}
\includegraphics[height=.25\textheight, angle =0]{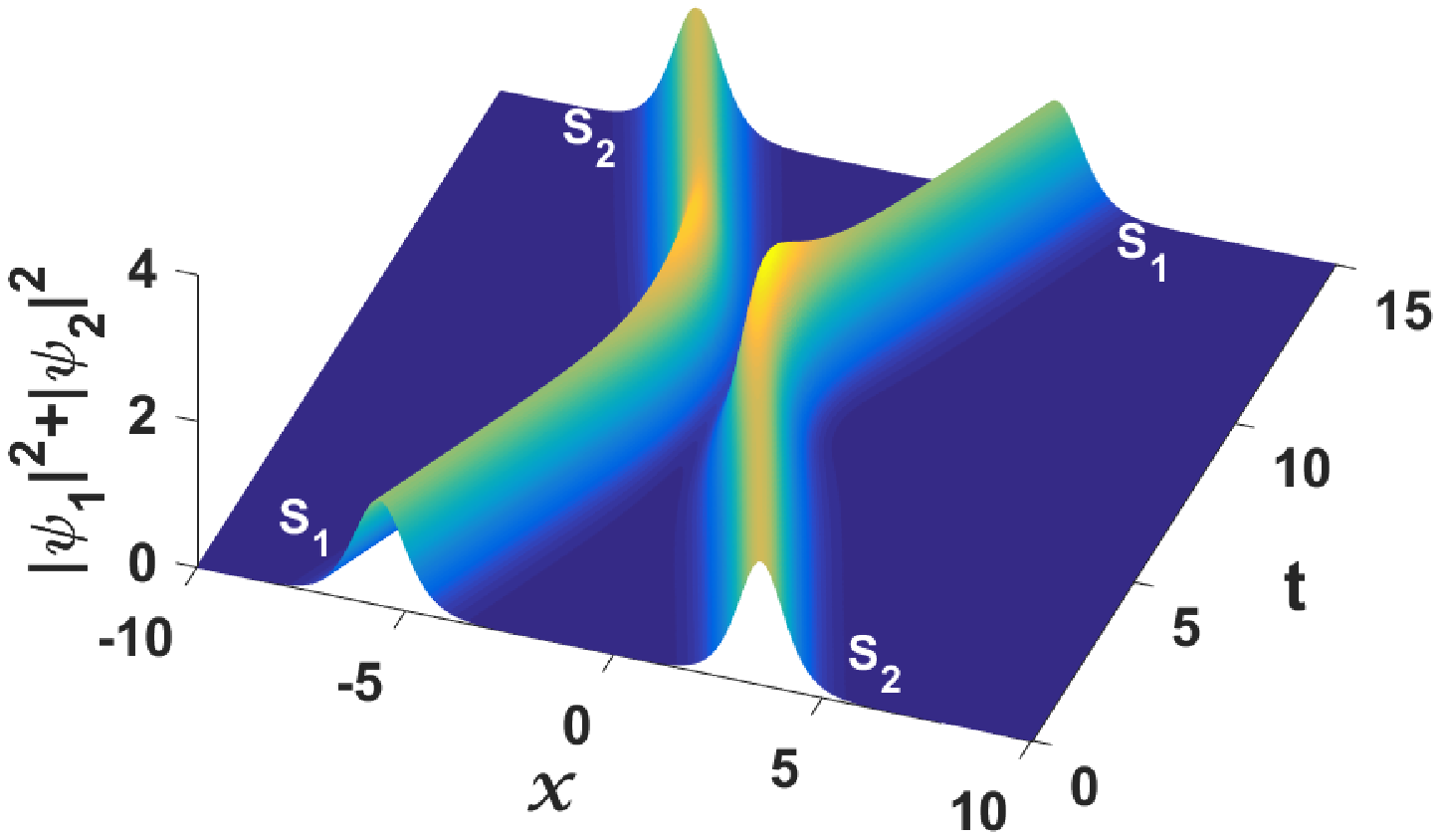}
\label{fig2g}
}
\end{center}
\caption{Panels (a)-(b): Elastic collision of oscillating solitons with $\alpha_1^{(1)}=\alpha_2^{(1)}=0.002$,
$\alpha_2^{(2)}=\alpha_1^{(2)}=0.005i$, $k_1=-1.4+0.5i$, and $k_2=1.6-0.5i$, $\beta=1$, and $\chi=2$.
Shape changing collision of oscillating solitons (c)-(d) with $\chi$ = 2 and other parameters are same as
given in Fig.~\ref{fig3}. Panel (e) shows the conservation of energy during shape changing soliton collision.}
\label{fig4}
\end{figure}

In this case, the two-soliton solution for constant nonlinearity parameter ($\beta(t)=1)$ and in the
absence of external potential ($V_{\mbox{ext}}(x,t)=0)$  corresponding to a homogeneous condensate is given by
 \bes\bea
\psi_1(x,t) &=& \frac{1}{D}\bigg(\cos(\chi t)G_1-i\sin(\chi t) G_2\bigg),\\
\psi_2(x,t) &= &\frac{1}{D}\bigg(-i\sin(\chi t)G_1+\cos(\chi t) G_2\bigg),
\label{twosoliton}
\eea\ees
 where $G_1$, $G_2$, and $D$ are defined in Eqs.~(\ref{g1} and \ref{d1}) in the appendix. Here, the form of $\eta_i,~ i=1,2,$
  as given
 in appendix is redefined as $\eta_i = k_i(x+ik_i (t/2))$. The top and middle row panels of Fig.~\ref{fig4}
 show elastic and energy sharing/shape changing collision of breathing
 solitons behaviour of system (\ref{lngp1}), respectively.
 These figures show that the Rabi coupling induces soliton oscillations which are spatially localized. Such breathing
 solitons can also be viewed as Ma-breathers. Particularly, panels \ref{fig2c}-\ref{fig2d} show
 elastic collision of oscillating solitons for the Rabi coupling parameter $\chi=2$. Figs.~\ref{fig2e}-\ref{fig2f} show
 the shape changing collision of bright solitons in  the presence of Rabi coupling. This Rabi coupling affects
 the switching dynamics significantly. We note that due to the Rabi effect the oscillation in soliton $S_1$ is
 completely
 suppressed before interaction while it reappears after interaction whereas soliton $S_2$
 exhibits
 oscillation before and after interaction. Another important effect
 one can notice is that in both the components the switching nature is same. That is, soliton $S_1$ gets enhanced
 after interaction. Meanwhile, the soliton $S_2$ which is completely suppressed before soliton collision in the
 Manakov case now reappears with significant amplitude and executes periodic oscillations in the $\psi_1$ component.
This is
 contrary to the collision scenario depicted in Fig.~\ref{fig3} where the Rabi coupling is absent.
 However, the total energy is conserved during shape changing collision as depicted in Fig.~\ref{fig2g} even in the presence of Rabi coupling.\\

\begin{figure}[H]
\begin{center}
\vspace{-0.2cm}
\mbox{\hspace{0.6cm}
\subfigure[][]{\hspace{-1.0cm}
\includegraphics[height=.25\textheight, angle =0]{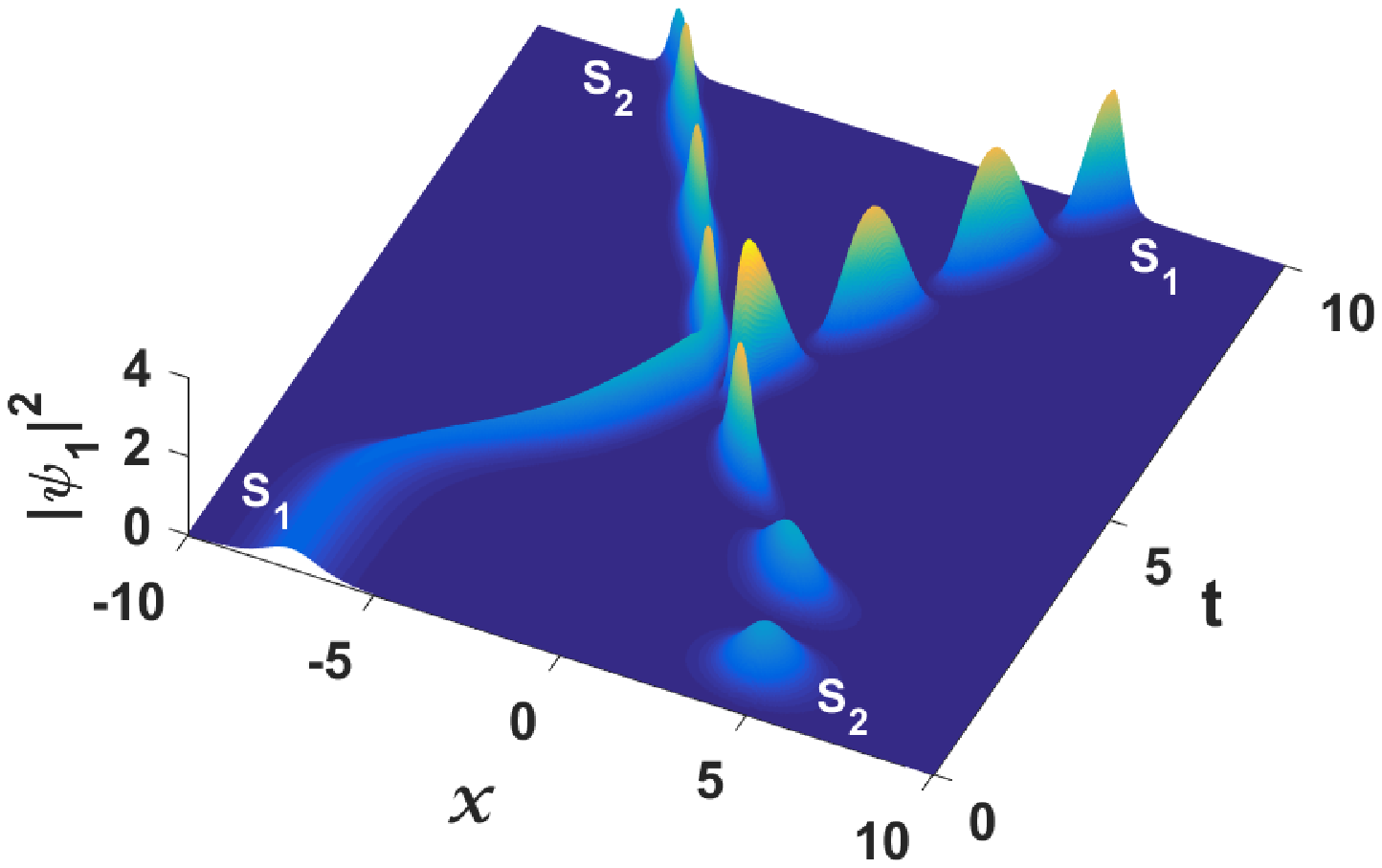}
\label{fig3c}
}
\subfigure[][]{\hspace{-0.2cm}
\includegraphics[height=.25\textheight, angle =0]{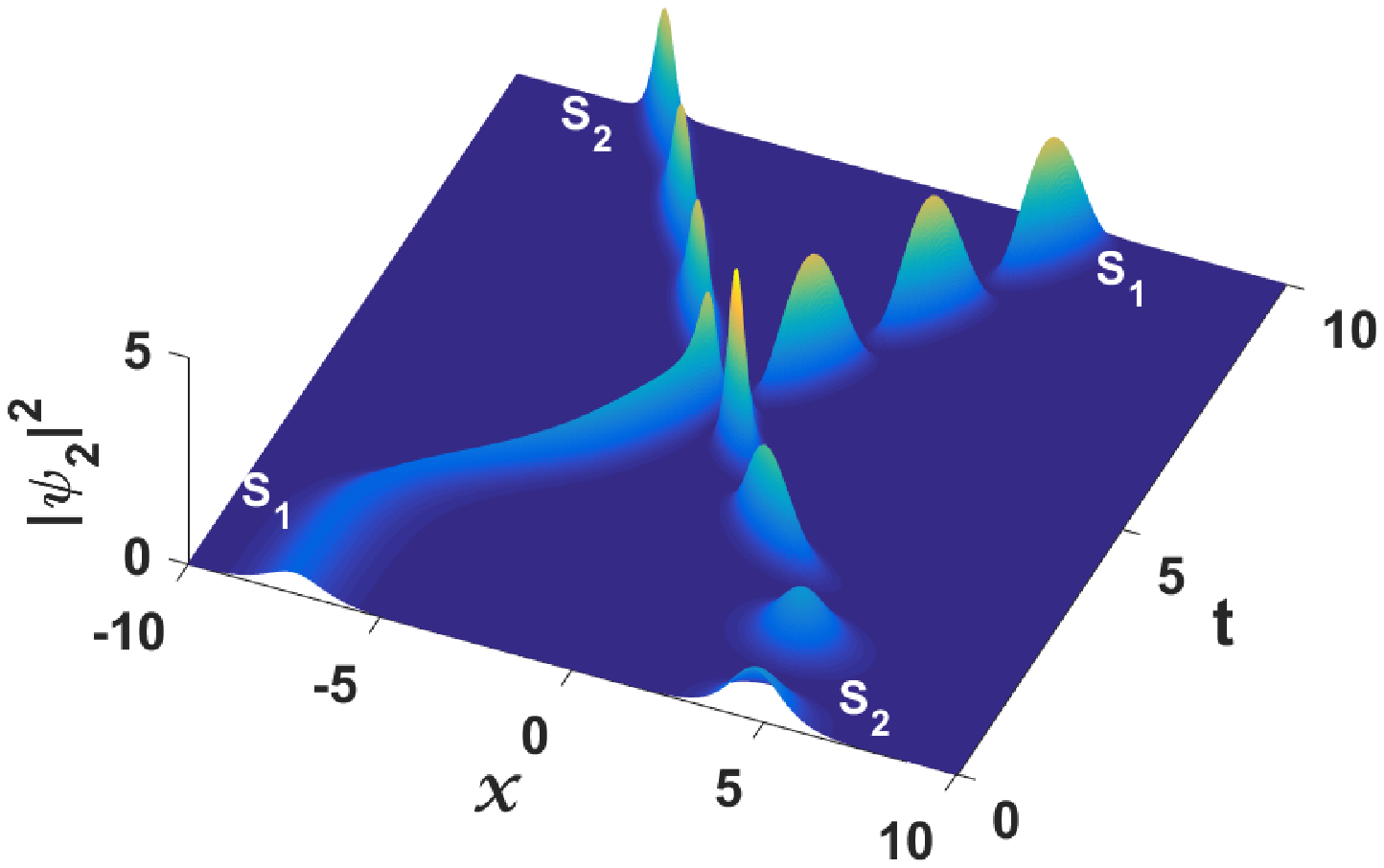}
\label{fig3d}
}
}
\mbox{\hspace{0.6cm}
\subfigure[][]{\hspace{-1.0cm}
\includegraphics[height=.25\textheight, angle =0]{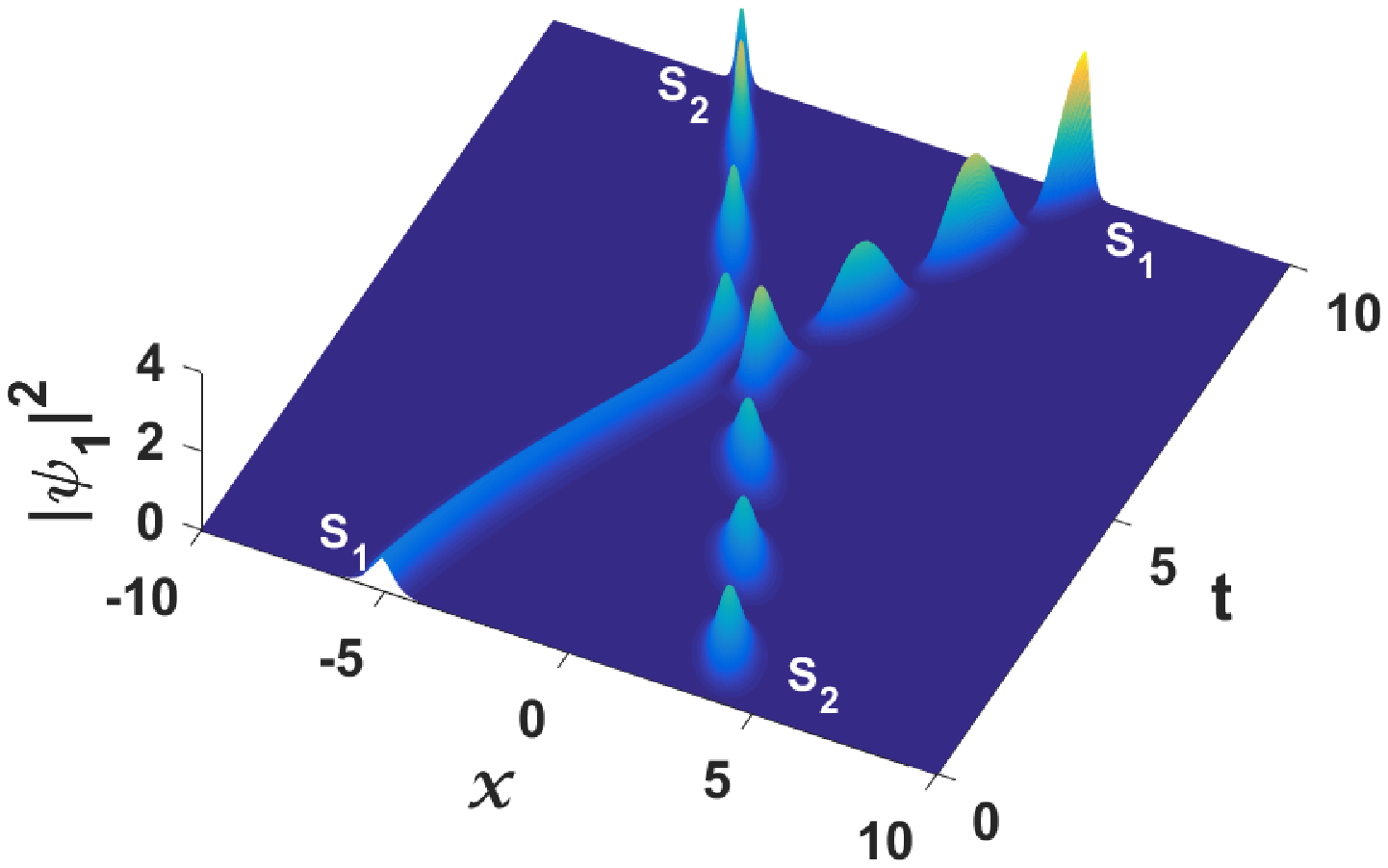}
\label{fig3e}
}
\subfigure[][]{\hspace{-0.2cm}
\includegraphics[height=.25\textheight, angle =0]{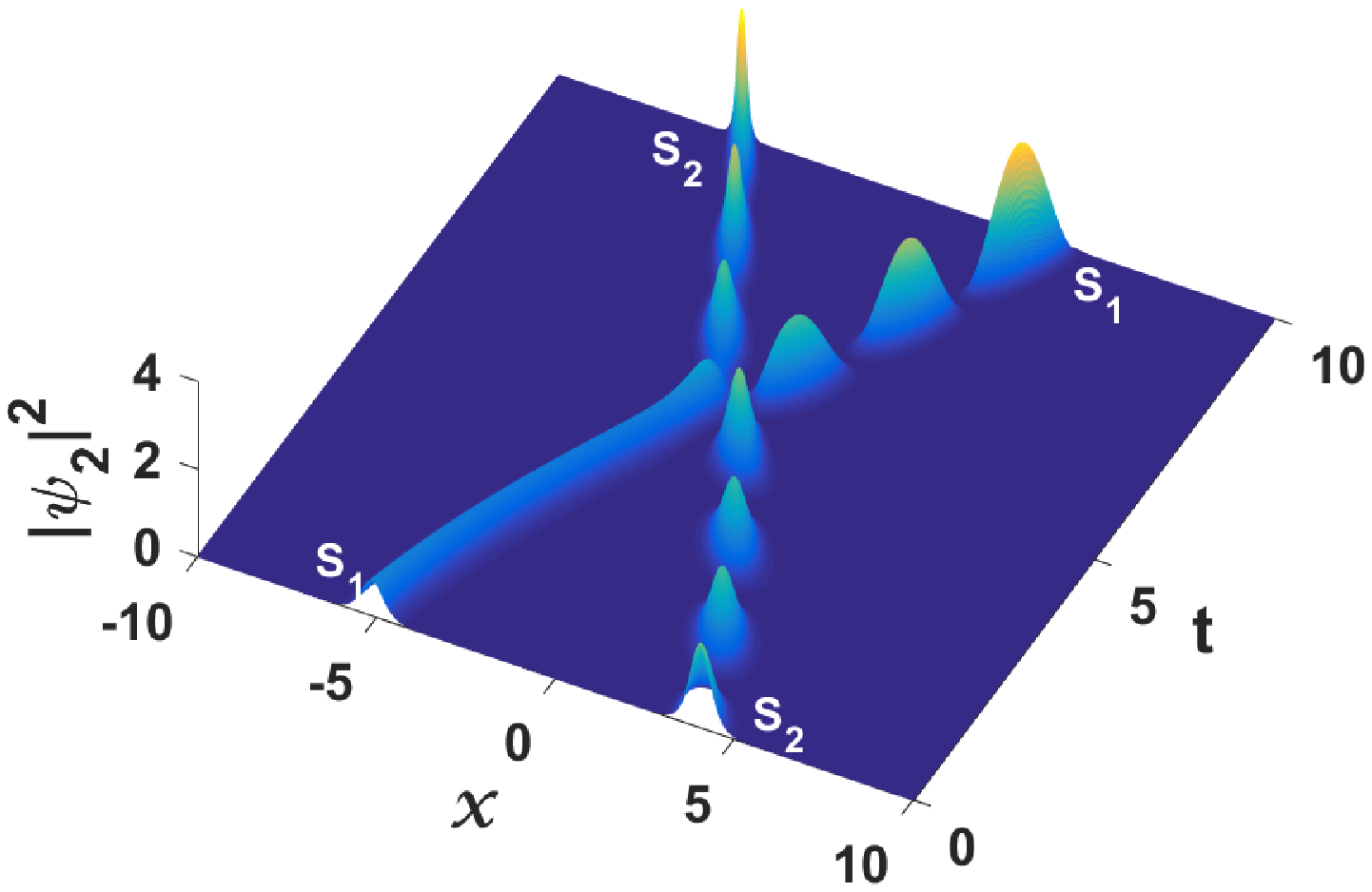}
\label{fig3f}
}
}
\end{center}
\caption{Panels (a)-(b): Shape changing collision of oscillating solitons correspond to values of
parameters of $\beta(t)=a_0+\tanh(\rho t+\delta)$, $\chi=2$, $\rho=1.2$, $\sigma_1=0.6$,
$\sigma_2=0$, $a_0=2$, and $\delta=-4$.
Panels (c)-(d): Shape changing collision with $\beta(t)=a_0+\cosh(\rho t+\delta)$
with parameters $\rho=0.35$, $\sigma_1=0.4$, $\delta=-1$, $\chi=2$, $a_0=2$, and $\sigma_2=0$.
 All other values are same as given in Fig.~\ref{fig3}.}
\label{fig5}
\end{figure}

{\bf (iii) Collision scenario in the presence of time-dependent nonlinearity coefficient $\beta(t)$ and
 external potential $V_{\mbox{ext}}(x,t)$}:

Next, we focus our study on soliton dynamics in the presence of the two time-varying nonlinearities discussed
in Sec.~IV and in the presence of an external potential.
The general two-soliton solution of the non-autonomous coupled GP system (\ref{lngp1}) can be written as
\bea
\left(
  \begin{array}{c}
    \psi_1 \\
    \psi_2 \\
  \end{array}
\right)= \frac{1}{D}\left(
  \begin{array}{cc}
    \cos(\chi t) & -i \sin(\chi t)\\
    -i  \sin(\chi t) &  \cos(\chi t) \\
  \end{array}
\right)
\left(
  \begin{array}{c}
    G_1 \\
    G_2 \\
  \end{array}
\right)\varepsilon(t) e^{i\phi(x,t)}.
\label{twosoliton}
\eea
Here, the co-ordinates $X$ and $T$ appear in the expressions of $G_1$ and $D$ (see appendix) and are redefined as follows:
$X$ = $\sqrt{2}\sigma_1\left(\beta(t)x-2 \sigma_2 \sigma_1^2\int \beta^2(t)dt\right)$,
$T$ = $\sigma_1^2\int \beta^2(t)dt$.
Figs.~\ref{fig3c}-\ref{fig3d} show the shape changing collision of two breathing solitons for
$\beta(t)$=$a_0+\tanh(\rho t+\delta)$.
The density of the breathing
soliton $S_1$ gets enhanced and $S_2$ is also enhanced after the collision in the $\psi_1$ component due to the
form of the kink nonlinearity. A similar behavior also takes place in the $\psi_2$ component. Thus, the switching
nature of
energy sharing collision in the autonomous system with Rabi coupling is affected by the presence
of the time dependent nonlinearity in the non-autonomous GP system (\ref{lngp1}) for this choice of time-dependent
nonlinearity and external potential. This clearly indicates that such type of kink-like
nonlinearity can be profitably used for soliton amplification by collision. One can note that the separation distance between the
solitons before and after interaction is also increased as compared
with Fig.~\ref{fig3}.
Finally, energy sharing collision of oscillating solitons in non-autonomous GP system (\ref{lngp1})
with $\beta(t)$=$a_0+\cosh(\rho t+\delta)$ is shown in Figs.~\ref{fig3e}-\ref{fig3f}.
 We observe that in this case also the energy sharing collision for the non-autonomous GP system (\ref{lngp1})
 with Rabi coupling is altered due to the nature of the nonlinearity coefficient.
 This nonlinearity can also be used advantageously for soliton amplification
 purpose. This collision can be viewed as interacting soliton cradles. Here there is a
 bending in the path of the colliding solitons. However suppression of oscillation in soliton $S_1$ is still
 preserved before interaction. The separation distance is left unaffected before and after collision as
 compared with Figs.~\ref{fig2e} and \ref{fig2f}.
\section{Stability of non-autonomous bright solitons}
\begin{figure}[H]
\begin{center}
\vspace{-0.2cm}
\mbox{\hspace{0.5cm}
\subfigure[][]{\hspace{0.1cm}
\includegraphics[height=.25\textheight, angle =0]{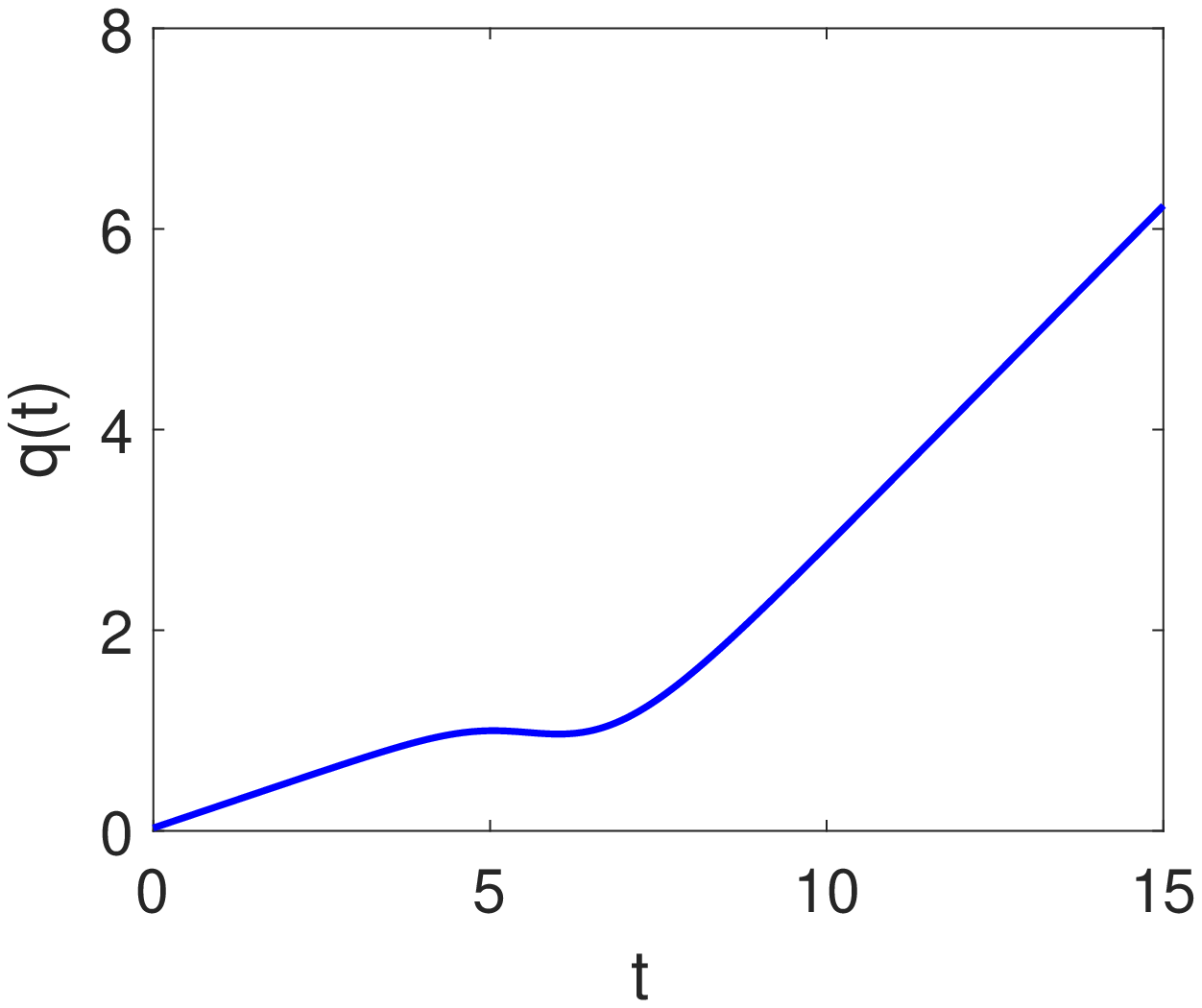}
\label{fig4a}
}
\subfigure[][]{\hspace{-0.2cm}
\includegraphics[height=.25\textheight, angle =0]{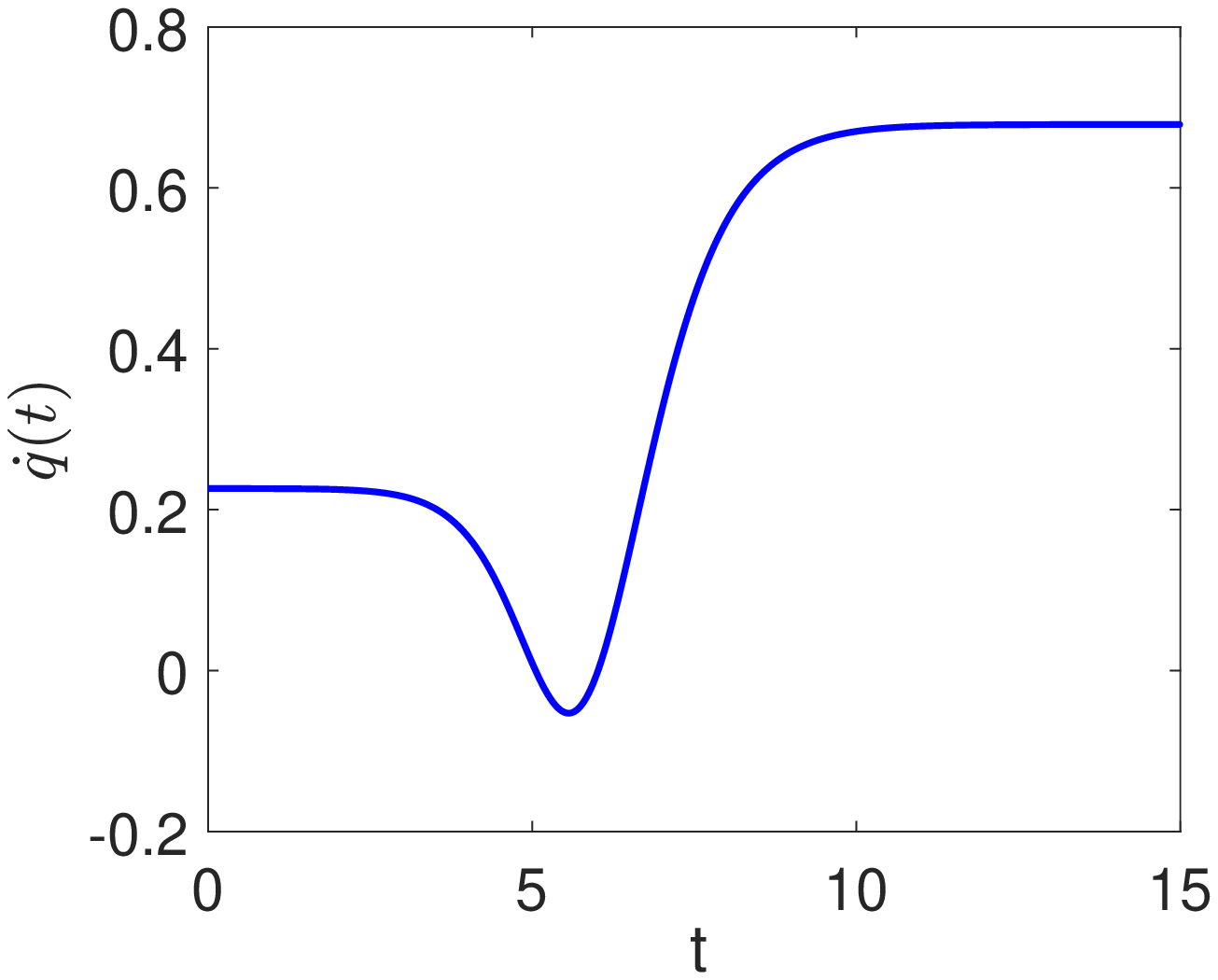}
\label{fig4b}
}
}
\mbox{\hspace{0.5cm}
\subfigure[][]{\hspace{0.1cm}
\includegraphics[height=.25\textheight, angle =0]{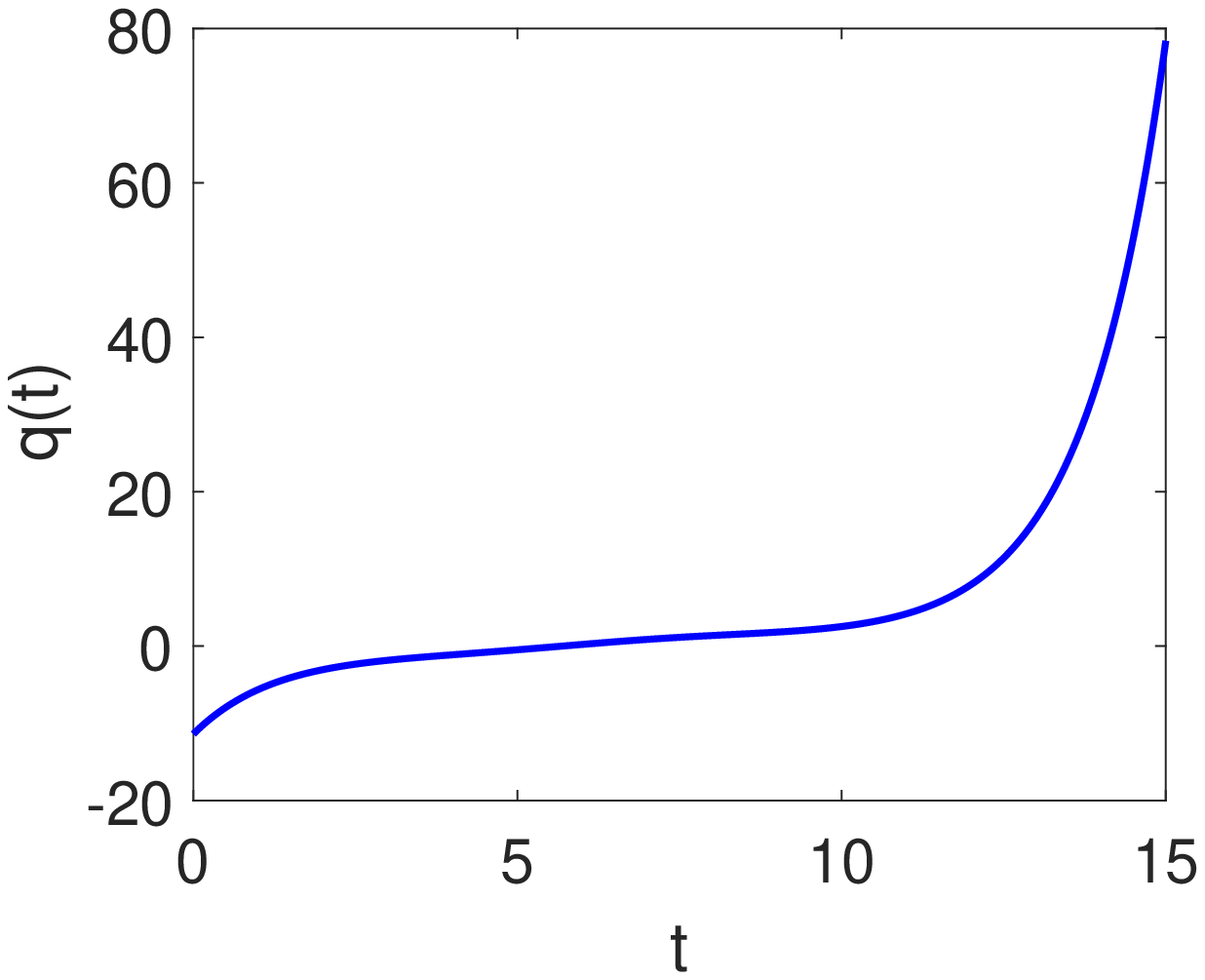}
\label{fig4c}
}
\subfigure[][]{\hspace{-0.2cm}
\includegraphics[height=.25\textheight, angle =0]{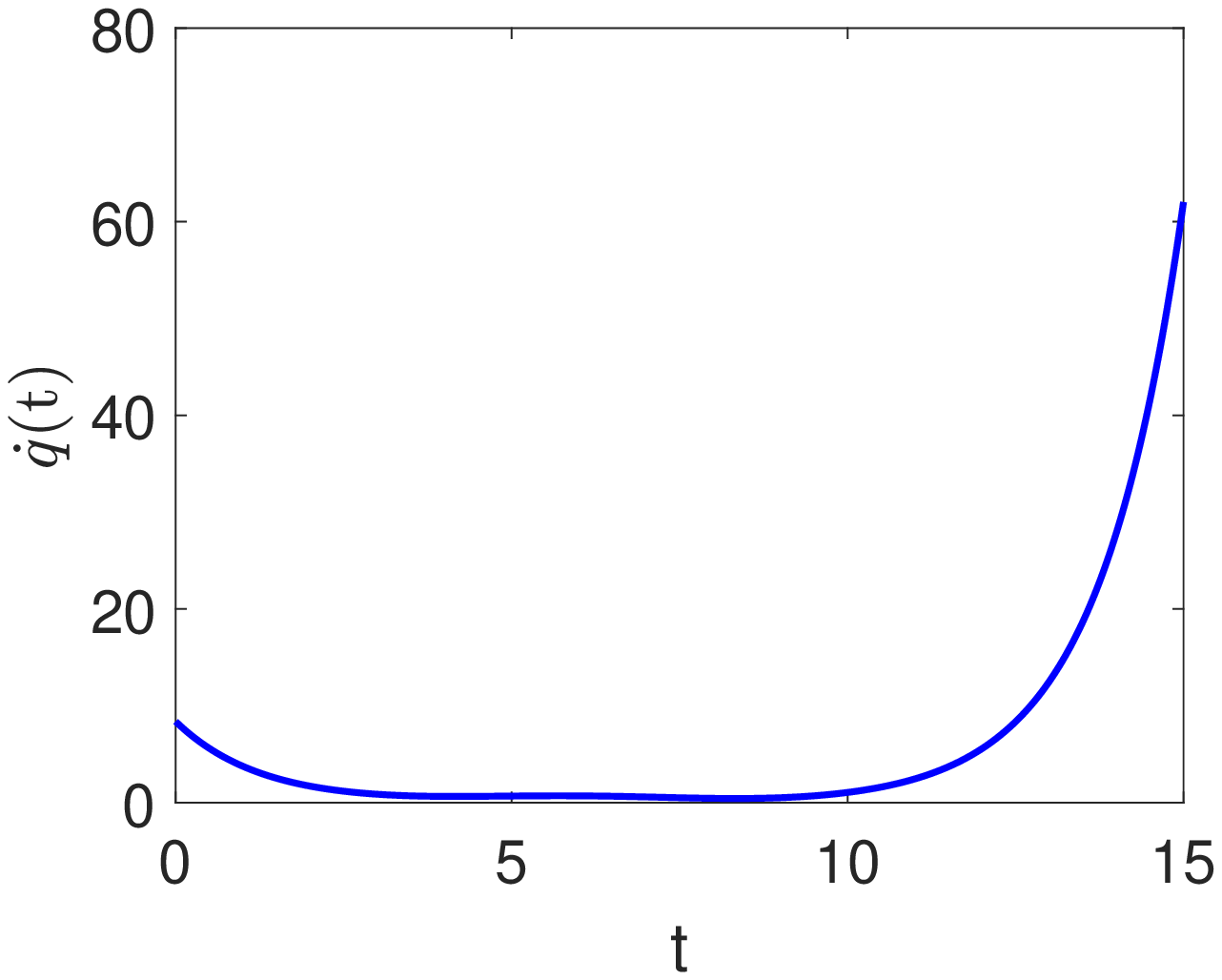}
\label{fig4d}
}
}
\end{center}
\caption{Panels (a)-(d) shows the position and velocity of the bright one soliton solution of system (\ref{lngp1}) with
$\beta(t)=2+\tanh(\rho t+\delta)$ and $\beta(t)=2+\cosh(\rho t+\delta)$, respectively.
The parameters are $k_{1R}=1$, $k_{1I}=0.2$, $\alpha_1^{(1)}=0.5$, $\alpha_1^{(2)}=0.2i$, $\sigma_1=0.8$,
$\sigma_2=0$, $\delta=-5$, $a_0=2$, and $\rho=0.8$.}
\label{fig6}
\end{figure}

Though the autonomous Manakov solitons are found to be stable \cite{bound,kannaprl}, it is not apparent that the obtained non-autonomous solutions are stable. So our next aim is to investigate the stability of the above discussed non-autonomous solitons.
In a recent interesting paper it has been shown that the sufficient and necessary condition for
soliton instability and stability is $dP(\upsilon)/d\upsilon<0$ and $dP(\upsilon)/d\upsilon>0$, respectively \cite{stability}.
Here, $P$ and $\upsilon$ are the normalised momentum and the soliton velocity.
In connection with the studies on soliton stability we need the explicit expressions for the norm,
soliton position and velocity and normalized momentum. For this purpose, here we calculate the following
conserved quantities of non-autonomous system (\ref{lngp1}).
Using the expression (\ref{norm}) and making use of (13) we find
the norm of the soliton is
\bea
N &= &\int_{-\infty}^{\infty}\rho~dx,\nonumber\\
&=&2\sigma_1^2\int_{-\infty}^{\infty}k_{1R}^2\beta(t) \mbox{sech}^2(k_{1R}\tilde{\omega}+R/2)dx\nonumber,\\
&=&2\sqrt{2}\sigma_1k_{1R}
\eea
which is time-independent.
By requiring the maximum of the condensate density $|\psi_j|^2,~j=1,2$, occurs at the
soliton position $x=q$, from (13), we get
\bes\bea
\sqrt{2}\sigma_1\left(\beta(t)q-2\sigma_2\sigma_1^2\int \beta^2(t)dt\right)=2k_{1I}\sigma_1^2\int \beta^2(t)dt-(R/2).
\eea
The resulting soliton position at which maximum condensate occurs is given by
\bea
q(t) &= &\frac{\zeta}{\sqrt{2}\sigma_1\beta(t)}.
\eea
By differentiating the above expression with respect to time `t' one can obtain the velocity of non-autonomous
soliton as
\bea
\dot{q}(t)& =&\frac{-\dot{\beta \zeta}}{\sqrt{2}\beta^2\sigma_1}+\sqrt{2}\sigma_1\beta(t)k_{1I}+2\sigma_1^2\sigma_2\beta(t),
\eea
where
\bea
\zeta=\left(2k_{1I}\sigma_1^2\int \beta^2(t)dt-\frac{R}{2k_{1R}}+2\sqrt{2}\sigma_1^3\sigma_2 \int \beta^2(t) dt\right).
\eea\ees
The top  panels [see Fig.~\ref{fig4a}-\ref{fig4b}] of Fig.~\ref{fig6} 
show the evolution of position and velocity of the non-autonomous bright
soliton for the nonlinearity parameter
$\beta(t)$ = $a_0+\tanh(\rho t+\delta)$. In this case, the soliton position increases linearly with respect to time  except in the jump region of the kink-nonlinearity and in this region it remains almost constant. For this kink nonlinearity, the soliton velocity  reaches a
negative minimum  and attains a constant maximum gradually. The bottom row panels
[see Fig.~\ref{fig4c}-\ref{fig4d}] of Fig.~\ref{fig6} show the position and velocity of the
soliton for the nonlinearity $\beta(t)$ = $a_0+\cosh(\rho t+\delta)$. In this scenario,
there is a shift in the soliton position
from negative to positive value as time evolves, whereas the soliton velocity initially takes a smaller
value and becomes zero for a significant time period followed by a steep
increase. Thus from Figs.~\ref{fig6}, it is
quite clear that
one can engineer the central position and velocity of the non-autonomous soliton by suitably
choosing the temporal modulations of the nonlinearity.
This arbitrariness of tuning the soliton position and velocity is not at all possible in its autonomous
counterpart.
The field momentum of soliton $p$ is obtained after substituting the one soliton solution (13)
in the following expression:
\bes\bea
p = \int_{-\infty}^{\infty}~j(x,t)~dx,
\label{momentum}
\eea
where the current density
\bea
\hspace{-0.8cm} j(x,t) = 2[- x \dot{\beta} +\sqrt{2}\sigma_1k_{1I}\beta^2(t)+2\sigma_1^2\sigma_2\beta^2(t)]k_{1R}^2\sigma_1^2 \mbox{sech}^2(k_{1R}\tilde{\omega}+R/2).
\eea\ees
 We get the final expression for the momentum current density after substituting $j(x,t)$ in Eq.~(\ref{momentum}) as
\bes\bea
p= 4 k_{1I} k_{1R}\beta(t)\sigma_1^2 +4\sqrt{2} k_{1R} \sigma_1^3 \sigma_2 \beta(t)- \frac{\dot{\beta}}{\beta^2}\bigg(4\sigma_1^2k_{1I}k_{1R}\int \beta^2(t)dt\nonumber\\
+4\sqrt{2}k_{1R}\sigma_1^3\sigma_2\int \beta^2(t)dt-R\bigg).
\eea
So the normalized momentum is given by
\bea
P = \frac{p}{N}=\frac{-\dot{\beta}}{\sqrt{2}\beta^2(t)\sigma_1}\left(2k_{1I}\sigma_1^2\int \beta^2(t) dt+2\sqrt{2}\sigma_1^3\sigma_2\int \beta^2(t) dt-\frac{R}{2k_{1R}}\right)\nonumber\\
+\beta\sigma_1\left(\sqrt{2}k_{1I}+2\sigma_1\sigma_2\right),
\eea
\bea
 P&=&\dot{q}(t).
\label{momentum1}
\eea\ees
From (\ref{momentum1}) we find $P(\upsilon)$ = $\upsilon$, where velocity $\upsilon=\dot{q}(t)$. Then we get
\bea
\frac{dP(v)}{dv}&=&1.
\eea
Hence the necessary condition  $dP(\upsilon)/d\upsilon >0$ for the stability of the soliton is fulfilled.
Let us compare the above expressions for position, velocity,
and momentum
 to the scalar NLS system.
Here, in vector NLS system the parameters $\alpha_1^{(1)}$ and $\alpha_1^{(2)}$ resulting
due to the vector nature of the system (\ref{lngp1}) influence the
position $q(t)$, velocity $\dot{q}(t)$ and hence the momentum $P(t)$ nontrivially.
\section{Conclusion}
We have studied the dynamics of
bright one- and two-solitons in 1D Rabi-coupled BEC with constant and time-dependent nonlinearity
coefficients. With the aid of  unitary and similarity/lens-type transformations, the 1D non-autonomous GP
system (\ref{lngp1}) is reduced to the standard integrable Manakov system. We present the bright one- and two-soliton solutions of the
Manakov system in the appendix. Then by making use of
these soliton solutions, the explicit soliton solutions of the non-autonomous GP system
(\ref{lngp1}) are constructed. The dynamics of bright solitons in the non-autonomous GP system is explored for two forms of
physically interesting nonlinearity coefficients, namely hyperbolic nonlinearities. From an application point of
view, the temporal modulations of the external potentials corresponding to these two choices of time dependent nonlinearities
are identified to have a close, rather almost same, resemblance with the superposed Hermite-Gaussian pulses which can be experimentally realized with modern day lasers. This
will open up a way in performing experiments on binary condensates to tune the potential to a desirable form. Specifically,
we show that in two-soliton case, the Rabi coupling produces breathers during two-soliton interaction. We
 also point out the
 interesting fact that due to the Rabi coupling the energy switching scenario in the Manakov system is altered preserving
 the total energy. The effect of kink-like nonlinearity is to result in a growth in the amplitude of the two colliding
 solitons after interaction with significant condensate compression. Also, the separation distance between the solitons, before interaction
 is increased as compared with the Manakov soliton interaction. Next, the effect of ``cosh" type nonlinearity results
 in a oscillating soliton collision with a complete suppression of oscillation in soliton $S_1$ before collision in both
 the components. The central position of the soliton also oscillates periodically during the collision.
  One can observe for this case, the separation distance between the solitons, before interaction remains the same
  as that of
  the Manakov solitons. Finally, for the first time to the best of our knowledge we have addressed analytically
  the stability of multicomponent non-autonomous bright solitons. Particulary, we have shown that the
  non-autonomous bright solitons are indeed stable as the rate of change of normalized momentum with respect to
  velocity is positive (i.e., $dP(v)/dv>0$).
We hope that the results of our study will be of use in the experimental realization of such solitons in
 binary BECs and will
 facilitate the understanding of the collisional properties of non-autonomous solitons in the Bose condensate mixtures.
\section*{Acknowledgements}
The work of T.K. is supported by Science and Engineering Research Board,
Department of Science and Technology (DST-SERB), Government of India,
in the form of a major research project (File No. EMR/2015/001408). F.G.M. thanks
Alexander von Humboldt-Foundation for travel support.

\appendix
\section{One- and two- soliton solutions of the integrable Manakov system (8)}
\setcounter{section}{1}
\subsection{Bright one-soliton solution}
The bright one-soliton solution of equation (8) reads as \cite{radha,kanna2003}
\bea
Q_j(X,T)=A_j k_{1R}~\mbox{sech}\bigg(k_{1R}\omega+\frac{R}{2}\bigg)e^{i\eta_{1I}}, \quad j=1,2,
\label{b-one}
\eea
where $\omega$ = $X-2k_{1I}T$, $\eta_{1I}$ = $k_{1I} X+(k_{1R}^2-k_{1I}^2)T$, $k_1=k_{1R}+ik_{1I}$,
 $R$= log$\left[(|\alpha_1^{(1)}|^2+|\alpha_1^{(2)}|^2)/(k_1+k_1^*)^2\right]$,
 $A_1=\frac{\alpha_1^{(1)}}{\sqrt{|\alpha_1^{(1)}|^2+|\alpha_1^{(2)}|^2}}$,
 $A_2=\frac{\alpha_1^{(2)}}{\sqrt{|\alpha_1^{(1)}|^2+|\alpha_1^{(2)}|^2}}$
 with $|A_1|^2+|A_2|^2=1$. Here, suffices $R$ and $I$ denote the real and imaginary parts.

\subsection{Bright two-soliton solution}
The bright two-soliton solution of equation (8) is given by \cite{radha,kanna2003}
\bea
Q_{j}(X,T) = \frac{G_j}{D}, \quad j=1,2,
\label{two-sol}
\eea
where
\bea
\hspace{-1cm}G_j = \alpha_1^{(j)}e^{\eta_{1}}+\alpha_2^{(j)}e^{\eta_{2}}+e^{\eta_{1}+\eta_{2}+\eta_{1}^*+\delta_{1j}}+e^{\eta_{1}+\eta_{2}+\eta_{2}^*+\delta_{2j}}, j=1,2,\label{g1} \\ \hspace{-1cm}D=1+e^{\eta_{1}+\eta_{1}^*+R_1}+e^{\eta_{1}+\eta_{2}^*+\delta_0}+e^{\eta_{2}+\eta_{1}^*+\delta_0^*}+e^{\eta_{2}+\eta_{2}^*+R_2}+
e^{\eta_{1}+\eta_{1}^*+\eta_{2}+\eta_{2}^*+R_3},
\label{d1}\eea
where, $\eta_i$ = $k_i(X+ik_i T)$, $e^{\delta_0}=\kappa_{12}/(k_1+k_2^*)$, $e^{R_1}=\kappa_{11}/(k_1+k_1^*)$,  $e^{R_2}=\kappa_{22}/(k_2+k_2^*)$, $e^{R_3}=|k_1-k_2|^2/(k_1+k_1^*)(k_2+k_2^*)|k_1+k_2^*|^2$, $e^{\delta_{1j}}=(k_1-k_2)(\alpha_1^{(j)}\kappa_{21}-\alpha_2^{(j)}\kappa_{11})/(k_1+k_1^*)(k_1^*+k_2)$,
$e^{\delta_{2j}}=(k_2-k_1)(\alpha_2^{(j)}\kappa_{12}-\alpha_1^{(j)}\kappa_{22})/(k_2+k_2^*)(k_1+k_2^*)$,
$e^{\delta_{2j}}=|k_1-k_2|^2(\kappa_{11}\kappa_{22}-\kappa_{12}\kappa_{21})/(k_2+k_2^*)(k_1+k_1^*)|k_1+k_2^*|^2$, and
$\kappa_{il}=\sum_{n=1}^2\alpha_i^{(n)}\alpha_l^{(n)^*}/(k_i+k_l^*)$, $i,l=1,2.$

\end{document}